\documentclass[PAPER, cernpreprint, coverpage=false, atlasdraft=false, texlive=2023, UKenglish, texmf, orcidlogo]{atlasdoc}

\usepackage{atlaspackage}
\usepackage{atlasbiblatex}
 
\usepackage{atlasphysics, atlasheavyion, atlasprocess, atlasparticle, atlasxref, atlasjetetmiss}
 
\usepackage{atlascontribute}

\graphicspath{{logos/}{figures/}}
\usepackage{ANA-HION-2021-12-PAPER-defs}
\usepackage{multirow}
\usepackage{scalerel}
\usepackage{etoolbox}
 
\usepackage{ANA-HION-2021-12-PAPER-defs}

\AtlasTitle{Production of $\varUpsilon(\text{nS})$ mesons in Pb+Pb and \textit{pp} collisions at 5.02~TeV}
 
\AtlasVersion{3.0}
 
\AtlasAbstract{
A measurement of the production of vector bottomonium states,
$\varUpsilon(\text{1S})$, $\varUpsilon(\text{2S})$, and
$\varUpsilon(\text{3S})$, in Pb+Pb and \textit{pp} collisions at a center-of-mass
energy per nucleon pair of 5.02~TeV is presented.
The data correspond to integrated luminosities of 1.38~$\text{nb}^{-1}$ of Pb+Pb
data collected in 2018, 0.44~$\text{nb}^{-1}$ of Pb+Pb data collected in 2015,
and 0.26~$\text{fb}^{-1}$ of \textit{pp} data collected in 2017 by the ATLAS detector
at the Large Hadron Collider.
The measurements are performed in the dimuon decay channel for
transverse momentum $p_\text{T}^{\mu\mu} < 30$~GeV, absolute rapidity
$|y^{\mu\mu}|<1.5$, and Pb+Pb event centrality 0--80\%.
The production rates of the three bottomonium states in Pb+Pb collisions are compared with those
in \textit{pp} collisions to extract the nuclear modification factors
as functions of event centrality, $p_\text{T}^{\mu\mu}$, and $|y^{\mu\mu}|$.
In addition, the suppression of the excited states
relative to the ground state
is studied. The results are compared
with theoretical model calculations.
}

\AtlasRefCode{HION-2021-12}
 
\PreprintIdNumber{CERN-EP-2022-045}

\arXivId{arXiv:2205.03042}

\AtlasJournal{Phys.\ Rev.\ C}
\AtlasJournalRef{Phys. Rev. C 107 (2023) 054912}
\AtlasDOI{10.1103/PhysRevC.107.054912}

\hypersetup{pdftitle={ATLAS document},pdfauthor={The ATLAS Collaboration}}
 
\begin{document}
 
\maketitle

\section{Introduction}
 
Quantum chromodynamics (QCD) predicts that at high temperatures and energy densities, hadronic matter undergoes a phase transition and turns into a state
of deconfined quarks and gluons known as quark--gluon plasma (QGP).
This state of matter is typically thought to be created in the collisions of two heavy nuclei at ultra-relativistic energies.
In such collisions, heavy-flavor quarks, especially charm and bottom, are produced
at an early stage in hard scattering processes and hence can
probe QGP over its full evolution.
 
Formation of the QGP and the consequent modification to the heavy-quark potential is expected to
lead to different quarkonium states dissolving at
different temperatures of the medium~\cite{Matsui:1986dk}.
This effect is known as sequential suppression~\cite{Karsch:1987pv}.
While the excited states are dissociated
just above the transition temperature $T_\mathrm{c} \approx 155$ MeV needed to form the QGP,
the ground states melt far above that value, creating a hierarchy in the
measured suppression of quarkonium states.
In particular, in Ref.~\cite{Digal:2001iu}, lattice calculations for the
temperature-dependent behavior of the heavy-quark potential in full QCD theory
were used to estimate
the
order
of the suppression steps as  functions of temperature and
energy density.
It was found that the \Ups{1} persists well above
$T_\mathrm{c}$, while \Ups{2} dissociates at about 1.1\,$T_\mathrm{c}$ and \Ups{3} cannot exist at
temperatures above $T_\mathrm{c}$.
Since then, quarkonium production and propagation through QGP have been extensively
studied theoretically. Comprehensive reviews can be found
in Refs.~\cite{THRev1, THRev2}.
 
Quarkonia dissociation in QGP can happen along with recombination of uncorrelated
heavy quarks~\cite{DuRapp, Greiner2018, Yao:2021}, which increases quarkonia
yields.
In this regard, it is interesting
to compare bottomonium (\Upsi, \Upsii, \Upsiii, $\chi_{\mathrm{b}}$, etc.)
to the charmonium family (\Jpsi, $\psi(2S)$, $\chi_{\mathrm{c}}$, etc.), since
recombination is expected to be much larger for the latter.
 
Experimentally, quarkonium suppression in nucleus--nucleus
collisions has been studied extensively for both the
bottomonia~\cite{PHENIX-Ups-2015, STAR-Ups-2014, CMS-HIN-16-023, CMS-HIN-16-008, ALICE-Ups-2014, ALICE-Ups-2021, STAR-UU, STAR-NEW}
and charmonia~\cite{SPS2, SPS1, PHENIX-Jpsi-2007, STAR-Jpsi-2017, HION-2016-07, CMS-HIN-14-005, ALICE-Jpsi-2017}
families at RHIC and LHC energies.
These measurements show strong suppression of quarkonia in nucleus--nucleus
collisions compared to \pp collisions, increasing
for more central events,
as well
as stronger suppression of the excited Upsilon states (\Upsii and \Upsiii) relative to
the ground state (\Upsi).
 
In this paper, \Ups{n} production in \pp and \PbPb collisions
at $\sqrt{s} = 5.02~\TeV$ per nucleon--nucleon pair is studied as a function of
transverse momentum ($p_\text{T}^{\mu\mu}$), rapidity, and \PbPb collision centrality.

\section{The ATLAS detector}
 
The ATLAS detector~\cite{PERF-2007-01} at the LHC covers nearly the entire
solid angle around the collision point.\footnote{ATLAS uses a right-handed coordinate system with its origin at the nominal interaction point (IP) in the center of the detector and the $z$-axis along the beam pipe. The $x$-axis points from the IP to the center of the LHC ring, and the $y$-axis points upward. Cylindrical coordinates $(r,\phi)$ are used in the transverse plane, $\phi$ being the azimuthal angle around the $z$-axis. The pseudorapidity is defined in terms of the polar angle $\theta$ as $\eta=-\ln\tan(\theta/2)$, and the rapidity is defined as $y = (1/2)[(E+p_z)/(E-p_z)]$.}
It consists of an inner tracking
detector surrounded by a thin superconducting solenoid, electromagnetic
and hadronic calorimeters, and a muon spectrometer incorporating three
large superconducting  air-core toroidal magnets.
 
The inner-detector system (ID) is immersed in a \SI{2}{\tesla} axial magnetic
field and provides charged-particle tracking in the range \(|\eta| < 2.5\).
The high-granularity silicon pixel detector covers the vertex region and
typically provides four measurements per track, the first hit normally being
in the insertable B-layer installed
before Run~2~\cite{ATLAS-TDR-19,PIX-2018-001}.
It is followed by the silicon microstrip tracker, which usually provides
eight measurements per track.
These silicon detectors are complemented by the
transition radiation tracker (TRT), which enables radially extended track
reconstruction up to \(|\eta| = 2.0\).

The muon spectrometer (MS) comprises separate trigger and high-precision tracking
chambers
measuring the deflection of muons in a magnetic field generated by the superconducting
air-core
toroids. The field integral of the toroids ranges between \num{2.0} and
\SI{6.0}{\tesla\metre} across most of the detectors.
A set of precision chambers covers the region \(|\eta| < 2.7\) with three layers of
monitored drift tubes, complemented by cathode strip chambers in the forward region,
where the background is highest.
Resistive plate chambers (RPCs) and thin gap chambers (TGCs) with a coarse position
resolution but a fast response time are used primarily to trigger on muons in the ranges
$|\eta| < 1.05$ and $1.05 < |\eta| < 2.4$ respectively.
 
The zero-degree calorimeters (ZDCs) are located symmetrically at $z = \pm 140$ m and
cover $|\eta| > 8.3$. The ZDCs use tungsten plates as absorbers, and quartz
rods sandwiched between the tungsten plates as the active medium.
In Pb+Pb collisions, the ZDCs primarily measure “spectator” neutrons that do not
interact hadronically when the incident nuclei collide.
 
Centrality in \PbPb collisions is determined by measuring the total transverse
energy deposited in a
liquid-argon forward calorimeter (FCal), which covers the pseudorapidity
range $3.1 < |\eta| < 4.9$.
The FCal is approximately 10 interaction lengths deep, and consists of three
modules: the first, with copper absorbers, is optimized for
electromagnetic measurements, while the other two, with tungsten absorbers,
are mainly sensitive to energy depositions associated with produced hadrons.
 
A two-level trigger system is used to select events of interest~\cite{TRIG-2016-01}.
The first-level (L1) trigger is implemented in hardware and uses a subset of detector information
to reduce the event rate to a design value of at most \SI{100}{\kHz}.
This is followed by the software-based high-level trigger (HLT), which reduces the event
rate to about 1--\SI{4}{\kHz}.
The L1 muon trigger requires coincidences between hits on different RPC or TGC planes, which
are used as a seed for the HLT algorithms. The HLT uses dedicated algorithms to incorporate
information from both the MS and the ID, achieving position and momentum resolution close
to that provided by the offline muon reconstruction, as shown in Ref.~\cite{TRIG-2016-01}.
 
An extensive software suite~\cite{ATL-SOFT-PUB-2021-001} is used in the reconstruction
and analysis of real and simulated data, in detector operations, and in the trigger
and data acquisition systems of the experiment.
The offline event selection required that events pass in-time pileup cuts based
on the ZDCs energy.
 
\section{Data selection and simulation samples}
 
The results presented in this paper were obtained using $pp$ data recorded
in 2017 at a
center-of-mass energy of $5.02~\TeV$ as well as
\PbPb data collected in 2015 and 2018
at $5.02~\TeV$ per nucleon--nucleon pair.
 
The integrated luminosity of the analyzed \pp collision samples is 0.26~\ifb. The \pp
events were collected using a dimuon trigger which requires at least two
spatially separated muon
candidates at L1, while both satisfy the criterion of $\pt^{\mu} > 4~\GeV$ in the HLT.
For the \PbPb analysis, the integrated luminosity is 0.44~\inb\ for
the 2015 data sample and 1.38~\inb\ for the 2018 data sample.
\PbPb events were collected using triggers which require at least one muon
with $\pt^{\mu} > 4$~\GeV\ at both L1 and the HLT, and at least
one additional
muon
satisfying $\pt^{\mu} > 4$~\GeV\
in the HLT,
without requiring matching to L1.
 
Muon pairs were required to fulfill the following criteria:
at least one reconstructed muon matching the HLT's dimuon trigger,
and both muons matching the HLT without an L1 trigger requirement;
both muons satisfy the \textit{Medium} identification criteria (described in
Ref.~\cite{PERF-2015-10}) without any requirements on TRT hits;
both muons are associated with the primary vertex
reconstructed using all of the tracks in each event;
the muon pair has a
dimuon mass $7.7 < m_{\mu\mu} < 12.3~\GeV$, dimuon transverse momentum
$|\pt^{\mu\mu}| < 30~\GeV$ and dimuon rapidity $|y^{\mu\mu}| < 1.5$;
the selected muon pair is re-fitted to a common vertex, with the
vertex fit quality satisfying $\chi^2 < 100$ and the significance of the transverse displacement of
the re-fitted vertex
relative to the primary vertex satisfying $|L_{xy}/\sigma(L_{xy})| < 3$, where
$\sigma(L_{xy})$ is the primary vertex resolution.
The dimuon rapidity requirement was
chosen to be $|y^{\mu\mu}| < 1.5$ because at more forward rapidities the
$\varUpsilon$ mass resolution starts to deteriorate quickly.
 
Monte Carlo (MC) simulation is used to study the \Ups{n} acceptance, the fit model used
in $\varUpsilon$ signal extraction,
and the closure of muon reconstruction and identification corrections
(due to residual biases associated with the yield correction procedure).
The unpolarized prompt $\varUpsilon(nS)$ in pp events were generated
with the CTEQ6L1~\cite{Pumplin:2002vw} parton distribution function
set in \PYTHIA[8]~\cite{Sjostrand:2014zea} with subsequent decay in muon pairs.
\PYTHIA[8] implements prompt \Ups{n} production sub-processes using the non-relativistic QCD Color Octet mechanism~\cite{NRQCD:1995}.
Prompt \Ups{n} production includes prompt production from the hard interactions, as well
as the radiative feed-down from $\chi_{b}\rightarrow \Ups{n} \gamma$ decays.
The production of nonprompt \Jpsi, used for determining per-muon corrections,
was simulated in \PYTHIA[8] by forcing $b\bar{b}$ production and
and retaining only events consistent with
\Jpsi decays.
The \PbPb MC sample was created
by overlaying  simulated \PYTHIA[8] \pp events with recorded minimum-bias \PbPb
events,
so that the ``data overlay'' simulation samples contain the same level of
underlying-event activity as is present in the \PbPb data.
The response of the ATLAS detector was simulated~\cite{SOFT-2010-01} using \GEANT~\cite{Agostinelli:2002hh}.
The MC events are reconstructed with the same algorithms as used in data.

\section{Analysis procedure}
 
\subsection{Centrality definition in \texorpdfstring{\PbPb}{Pb+Pb}}
 
The transverse energy measured in the forward calorimeter, \sumETFcal,
in minimum bias events, is used to estimate the degree of overlap between the two colliding Pb nuclei.
Each centrality class corresponds to a fixed percentile in the \sumETFcal distribution of
minimum-bias events using the procedure described in
Ref.~\cite{HION-2016-07}, where a table containing exact centrality bin definitions can be found.
A Monte Carlo Glauber-based model~\cite{Glauber} is used to calculate the mean number of
participant nucleons, \avgNpart, and the mean nuclear overlap function, $\avgTAA$,
for each centrality class.
 
\subsection{Corrections to raw invariant mass distributions}
 
Before the dimuon invariant mass distributions are fit to extract Upsilon yields, each candidate dimuon pair is corrected with a weight that accounts for the \Ups{n} dimuon acceptance, trigger efficiency, and reconstruction efficiency.

The kinematic acceptance $\mathcal{A}$ is defined as the probability that both muons
from $\varUpsilon \rightarrow \mu^{+}\mu^{-}$ decay pass the fiducial selection
($p_\mathrm{T}^{\mu} > 4$~\GeV\ and $|\eta^{\mu}|<2.4$). The kinematic acceptance
is calculated from a generator-level simulation separately for
different \Upsins states as described in Ref.~\cite{BPHY-2011-01}.
The acceptance correction also accounts for final-state radiation from one
or both of the decay muons.
In principle, the acceptance could depend on the spin-alignment of the \Upsins.
In this analysis, \Upsins mesons are assumed to be produced unpolarized, following the
previous measurements
in \pp collisions~\cite{CMS-BPH-11-023,CMS-BPH-13-003,LHCb_Jpsi_polarization}.
These measurements are consistent with no $\varUpsilon$ polarization, but have
large uncertainties. No extra systematic uncertainty
due to $\varUpsilon$ polarization is added in this study.
 
The dimuon reconstruction efficiency, $\varepsilon_\mathrm{reco}(\mu_1\mu_2)$,
is determined as the product of two single-muon reconstruction efficiencies.
The single-muon reconstruction efficiency is factorized into ID track reconstruction
efficiency and MS reconstruction efficiency.
For \pp collisions, the values of the reconstruction efficiency are obtained from
the $\Jmm$ \PYTHIA[8] simulation, and additional data-to-MC efficiency scale
factors are derived using a $\Jmm$ tag-and-probe method~\cite{PERF-2015-10}
using \pp data
to account for residual differences between data and simulation.
The same $\Jmm$ tag-and-probe method is employed to measure the muon reconstruction
efficiency in \PbPb collisions. The ID reconstruction efficiency is obtained directly
from \PbPb data, with a requirement on the transverse displacement
of the \Jpsi vertex to suppress potential biases from displaced muons. The MS reconstruction efficiency is obtained from the
$\Jmm$ \PYTHIA[8] simulation overlaid with minimum-bias \PbPb data, and additional
data-to-MC scale factors are determined in \PbPb data to account for the small
difference between data and simulation.
The ID reconstruction efficiency is found to be larger than $99\%$ in both \pp
and \PbPb collisions, with a weak centrality dependence at low $\pt^{\mu}$ in the latter case.
The MS reconstruction efficiency at $\pt^\mu = 4~\GeV$ is about $65\%$ in the
barrel region ($|\eta_\mu| < 1.05$) and $75\%$ in the endcap
region ($1.05<|\eta_\mu| < 2.4$), and the MS efficiency increases with $\pt^\mu$
and saturates at $95\%$ around $\pt^\mu = 7~\GeV$ in both the barrel and
endcap regions.
Due to the absorption of most hadronic activity in the ATLAS calorimeters, the MS reconstruction efficiency has no centrality dependence.
For a given muon pair, the dimuon trigger efficiency in \pp collisions is
factorized as the product of two single-muon efficiencies.
The dimuon trigger efficiency in Pb+Pb collisions is the combined efficiency of
either of the two muons matching HLT trigger
while the other one
matching HLT trigger without L1 trigger requirement.
 
The single-muon trigger efficiency is determined in data and simulation using
a $\Jmm$ tag-and-probe method similar to that used for the muon reconstruction
efficiency.
Two different muon-trigger logic schemes are used in this analysis: 1) a full-chain
muon trigger, which requires the formation of an L1 muon candidate that is
subsequently confirmed in the HLT, and 2) a full-scan muon trigger, which is only
performed in the HLT via a muon candidate search of the full MS system without
any requirement at L1.
The values of the full-chain muon trigger efficiency are determined from
MC simulation, with data-to-MC scale factors determined from the \pp data to
take into account
the difference between data and simulation. The same values
are used for \pp and \PbPb collisions, except that
an additional centrality-dependent correction is applied to \PbPb data.
The centrality-dependent correction is determined as the ratio of the trigger
efficiency measured in \PbPb collisions as a function of centrality to
the trigger efficiency in \pp data.
The full-chain muon trigger efficiency plateau value is found to be $70\%$ in the barrel
and $90\%$ in the endcaps.
The centrality-dependent correction factor is about $90\%$ ($100\%$)
for the 0--10\% (60--80\%) centrality range.
The full-scan muon trigger is only used in \PbPb collisions and its efficiency
is determined directly from \PbPb data, using the tag-and-probe method.
The full-scan trigger plateau is found to be $90\%$ in both the barrel and endcap regions.
The dimuon trigger efficiency in \pp collisions is factorized as the product of two full-chain
muon trigger efficiencies, and in \PbPb collisions the factorization form consisting of full-chain
and full-scan muon trigger efficiencies, as detailed in Ref.~\cite{HION-2016-07}, is used.

\subsection{Upsilon signal extraction}
 
Upsilon states are reconstructed in the $\mu^+\mu^-$ decay channel and their
yields are determined via unbinned maximum-likelihood fits to the weighted dimuon
invariant mass
distributions, following the same procedure for both \pp and \PbPb data.
Each of the three \Upsins state signal shapes is described by a sum of
Crystal Ball~\cite{paper:CB} and Gaussian functions.
 
The probability distribution function for the fit is defined as a normalized sum of three $\varUpsilon$ signal components and a background component as
\begin{equation*}
\text{pdf}(m_{\mu\mu}) = N_{\Upsi}f_{\Upsi}(m_{\mu\mu}) + N_{\Upsii}f_{\Upsii}(m_{\mu\mu}) + N_{\Upsiii}f_{\Upsiii}(m_{\mu\mu}) + N_{\mathrm{bkg}}f_{\mathrm{bkg}}(m_{\mu\mu}),
\end{equation*}
 
where
$f_{\Upsins}(m_{\mu\mu}) = \omega  F_\text{G}(m_{\mu\mu};M_{\mathrm{nS}},\sigma_{\mathrm{nS}})+(1-\omega)F_\text{CB}(m_{\mu\mu};M_{\mathrm{nS}},1.7\sigma_{\mathrm{nS}},\alpha,n)$;
$N_{\Upsins}$ and $N_{\mathrm{bkg}}$ are the Upsilon and background
yields, respectively; and $F_\text{G}$ and $F_\text{CB}$ are Gaussian and Crystal Ball functions, respectively, with $\omega$ representing relative weight of the Gaussian
function.
The quantities $M_{\mathrm{nS}}$ and $\sigma_{\mathrm{nS}}$ are the Gaussian function's mean and width for each
Upsilon state, and $\alpha$ and $n$ are $F_\text{CB}$ tail parameters.
The line shape parameters are assumed to be the same for all three $\varUpsilon$ states since it
is mostly determined by detector effects.
The background shape, $f_{\mathrm{bkg}}(m_{\mu\mu})$, is represented by a
second-order polynomial
at high $\pt^{\mu\mu}$ ($\pt^{\mu\mu}>6$~\GeV) and as a product of an error function and an exponential
function at low $\pt^{\mu\mu}$.
 
The determination of the yield corrected for acceptance and efficiencies, $N_{\Upsins}^\mathrm{corr}$, proceeds in several steps.
First, a resonance-dependent weight, $w_{\mathrm{total}}(\Upsins)$, is determined
for each selected dimuon candidate as
 
\begin{equation*}
w_{\mathrm{total}}(\Upsins) = \frac{1}{\mathcal{A}(\Upsins) \varepsilon_\mathrm{reco}(\mu_1\mu_2) \varepsilon_\mathrm{trig}(\mu_1\mu_2) \varepsilon_\mathrm{pvAsso}(\mu_1\mu_2)} ,
\end{equation*}
where $\mathcal{A}(\Upsins)$ is the acceptance for $\Upsins\to\mu^+\mu^-$ decay,
$\varepsilon_{\mathrm{reco}}$ is the muon reconstruction efficiency, $\varepsilon_{\mathrm{trig}}$ is the trigger efficiency, and
$\varepsilon_\mathrm{pvAsso}$ is the efficiency related to the primary-vertex association.
Next, an unbinned maximum-likelihood fit to the weighted dimuon
invariant mass distribution ($m_{\mu\mu}$) is performed to extract the $\Upsins$ yields.
Three fits with an acceptance value corresponding to each state are performed to
extract the yields for the three $\varUpsilon$ states.
 
The mean and Gaussian width of the $\Upsi$ signal, $M_{1S}$ and $\sigma_{1S}$,
are left
unconstrained
in each $\pt^{\mu\mu}$, $|y^{\mu\mu}|$ and centrality range, while the means and
widths of $\Upsii$ and $\Upsiii$ in the same range are fixed to \Upsi
parameters scaled by the respective PDG~\cite{PDG-ptep} mass ratios.
The widths of the CB functions are set by scaling the corresponding Gaussian widths
by a constant factor, 1.7, which was determined in a previous
analysis~\cite{HION-2014-05} and validated with MC studies.
The relative weight of the Gaussian and CB functions, $\omega$, is not
constrained, and has the same value for all three Upsilon states.
For the \pp collision analysis, the CB function parameter $\alpha$,
which defines the point at which the low-mass tail transitions from a Gaussian shape to a
power-law shape, was fixed to the value obtained from \Upsi MC samples,
while $n$, which describes the shape of the tail,
was a free parameter.
For Pb+Pb collisions, both $\alpha$ and $n$ were fixed to the
values from the fit for \pp collisions, for each kinematic selection.
The nominal signal fit model described
above was validated by fitting the signal MC samples in various $\pt^{\mu\mu}$,
$|y^{\mu\mu}|$ and centrality intervals.
 
The background
functional form
varies with dimuon $\pt^{\mu\mu}$.
For $p_\mathrm{T}^{\mu\mu}>6$~\GeV\ the background is parameterized as a
second-order polynomial. However, for lower $p_\mathrm{T}^{\mu\mu}$
($p_\mathrm{T}^{\mu\mu}<6$~GeV),
and for the integrated over $\pt^{\mu\mu}$ fits,
in order to describe the
turn-on behavior of the dimuon acceptance caused by the single-muon
transverse momentum requirement, an error function
multiplied by an exponential function is used.
The background model parameters are initialized using a background-enriched
sample,
which consists of events with same-sign muon pairs, as well as a control sample consisting primarily of muons from b-hadron decays.
The control sample is produced by requiring at least one of the two muons to satisfy
$|\dzero|/\sigma_{\dzero} > 2$ or $|\zzsth| > 0.2$~mm, where $\dzero$ and
$z_0$ are the distances of closest approach of the muon to the primary vertex in
the plane perpendicular to the beam and in the beam direction, respectively.
After using the background distributions to determine the relevant parametrizations, the full model including signal and background contributions is used to fit the data, with the slope of the exponential function allowed to float.
 
Figure~\ref{fig:inclusivefit} shows an example of the fit to the dimuon mass plots
for \pp (left) and 0--80\% centrality \PbPb (right) collisions for the inclusive
$\pt^{\mu\mu}$ and $|y^{\mu\mu}|$ selection. The lower panels show the pull
distribution, which represents the distance between data points and fit function normalized by the data points statistical uncertainty:
$(\mathrm{Data}-\mathrm{Fit})/\sigma$(Data). The goodness of the fit is assessed by calculating the reduced chisquare, $\chi^{2}/\mathrm{NDF}$,  summing the squared deviations of the data points from the fit (weighted by the inverse errors) and then dividing by the number of degrees of freedom (NDF) in the fit. Typically, $\chi^{2}/\mathrm{NDF}$ for the fits varies from ${\approx}2.5$ to ${\approx}1$
for \pp and from ${\approx}2$ to ${\approx}1$ for Pb+Pb,  indicating that the data
and model agree within statistical uncertainties.
The extracted values of $\chi^{2}/\mathrm{NDF}$ decrease with $\pt^{\mu\mu}$, indicating that
the fit quality improves with $\pt^{\mu\mu}$.
Some relatively large $\chi^{2}/\mathrm{NDF}$ values are also found and can be explained by deviations
of the functions used for background description from the actual background
at the edges of the fit range.
 
\begin{figure}[htb]
\begin{center}
\includegraphics[width=\fourfigsize]{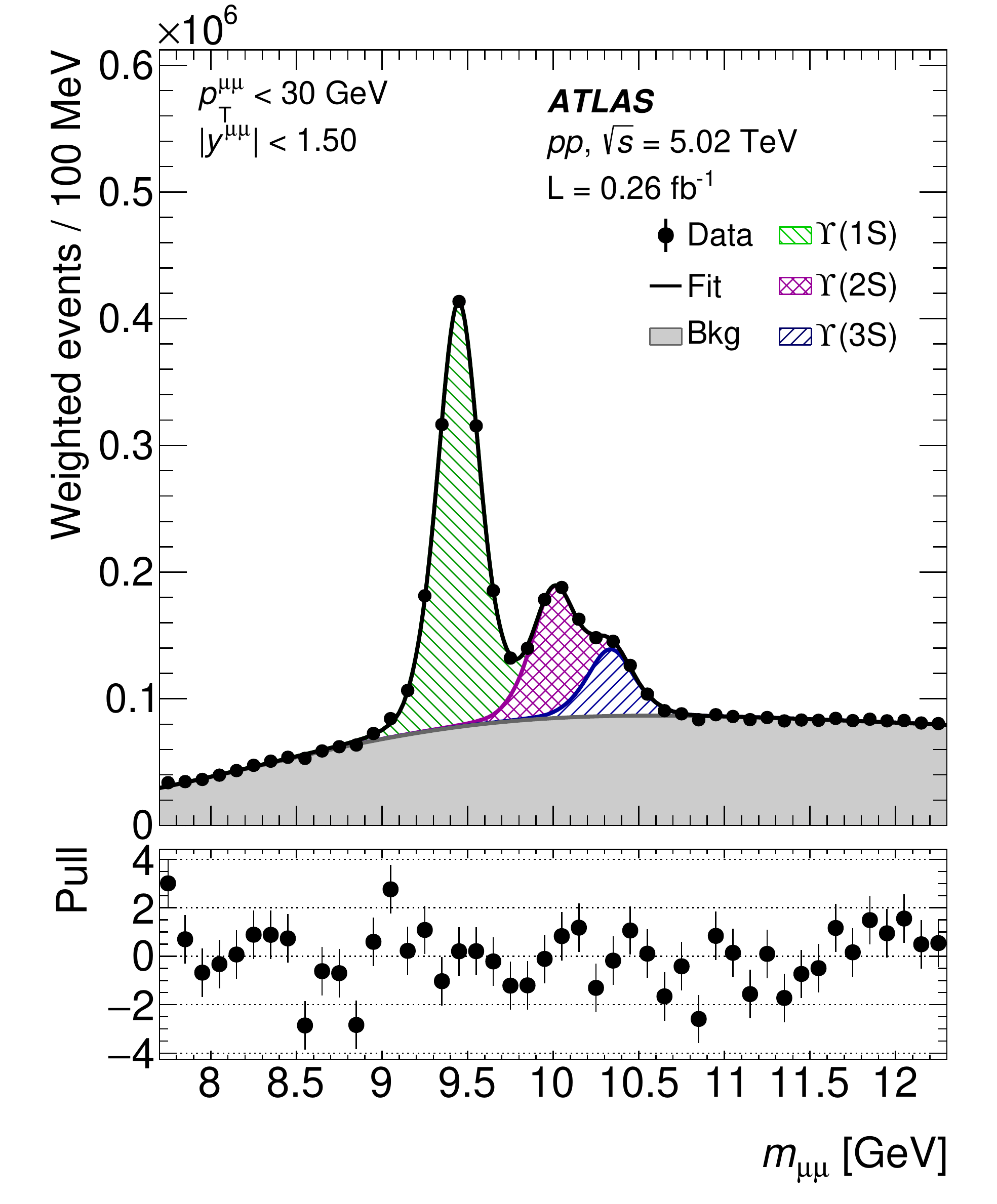}
\includegraphics[width=\fourfigsize]{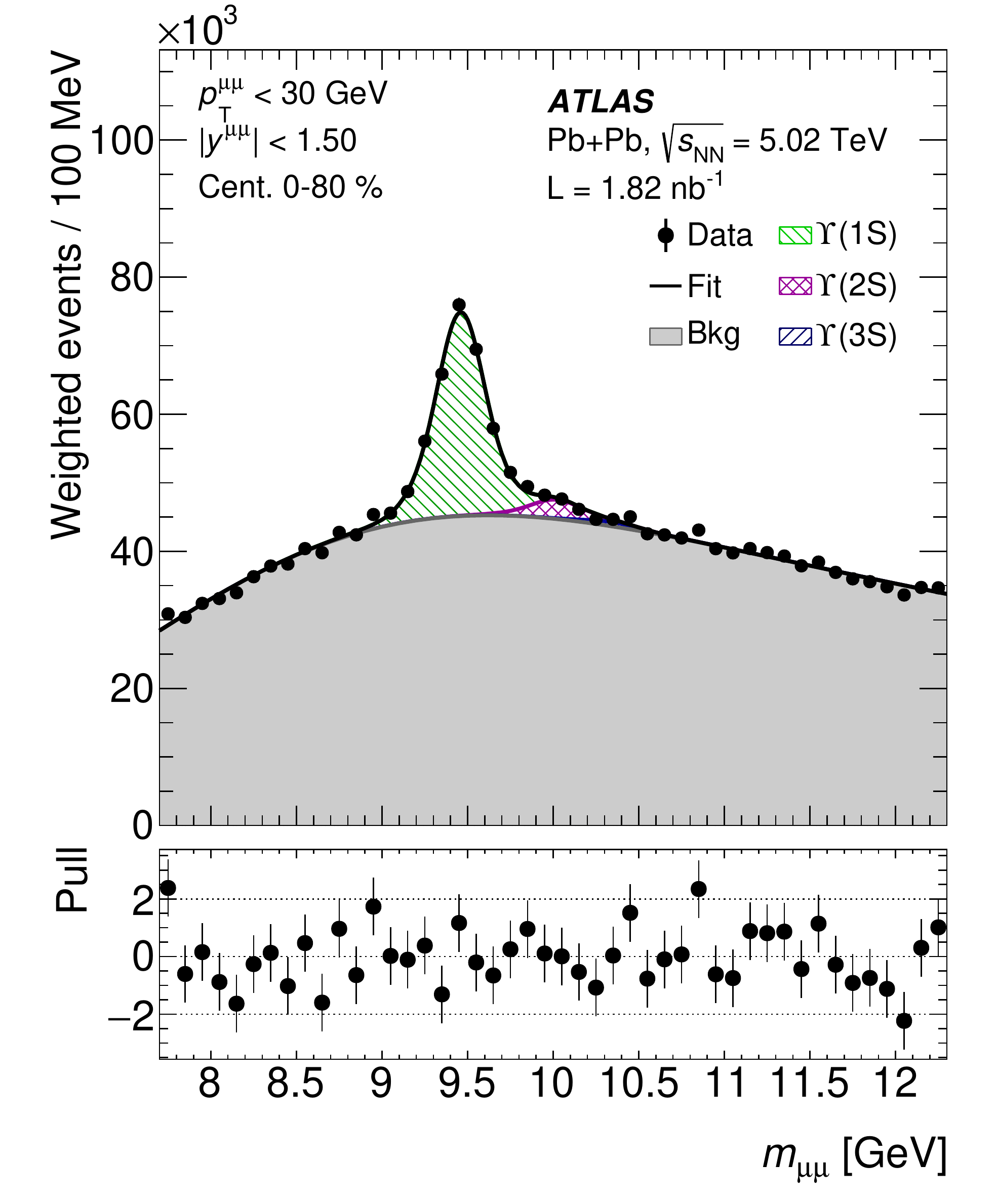}
\caption{Dimuon invariant mass distributions with the fit results for \pp (left) and \PbPb (right) collisions at 5.02~\TeV. The various curves are explained in the legend. Pull distributions are shown in the lower panels. For this selection, $\chi^{2}/\mathrm{NDF}$ is 2.7 for \pp and 1.3 for Pb+Pb.}
\label{fig:inclusivefit}
\end{center}
\end{figure}

\subsection{Systematic uncertainties}
 
The main sources of point-by-point uncorreleated systematic uncertainties
pertain to the corrections for
muon reconstruction and trigger efficiencies, and the yield extraction.
For the lowest $\pt^{\mu\mu}$ range,
the final-state
radiation correction is also a significant contribution.
Other, subdominant sources of systematic uncertainty considered in
this study are the primary-vertex association uncertainty and bin migration
due to momentum resolution.
Finally, the Upsilon states are assumed to be unpolarized and no extra systematic uncertainty
is assigned to cover this assumption.

The systematic uncertainty in the MS reconstruction efficiency is dominated by
the uncertainty in the data-to-MC scale-factor determination.
The scale-factor uncertainty is evaluated following Ref.~\cite{PERF-2015-10}
by changing the tag-muon selection criteria and varying the line-shapes in the efficiency
extraction fit procedure in the data.
The uncertainty related to the ID reconstruction efficiency, which is close to 1,
is estimated by comparing the results while varying this efficiency in both up and down directions by one standard deviation.
 
The systematic uncertainty in the muon trigger efficiency is also dominated by
the tag-and-probe efficiency determination procedure.
For \PbPb collisions, an additional systematic uncertainty associated with the
centrality-dependent correction is included.
This uncertainty is evaluated by comparing the centrality-dependence-corrected \PbPb
efficiency with the \pp efficiency as a function of $p_{T}^{\mu}$.
Individual variations
described above are added in quadrature
to form the total systematic uncertainty of the efficiency corrections.
 
The sensitivity of the signal extraction to the choice of a particular fit model
is evaluated by varying the line-shape of each fit component. The maximum
variation between the recalculated values and the central value is used to estimate
their uncertainty.
Eight variations are
considered, and these can be categorized into three groups: signal resolution
(width of the peak), shape of the final-state radiation tail,
and background shape.
The final uncertainty from the fit model is obtained by calculating the difference
between the maximum and the minimum yield from the eight line-shape variations and dividing
by $\sqrt{12}$, assuming a flat distribution.

The uncertainty in the $\varUpsilon$ acceptance due to the final-state radiation correction
was calculated by comparing
the result of a fully simulated acceptance calculation with that obtained using
an MC sample designed for a high-precision determination of the acceptance.
The signal Monte Carlo samples were processed with a fast simulation~\cite{SOFT-2010-01} which relies on a parameterization of the
calorimeter response~\cite{ATL-PHYS-PUB-2010-013}.
This uncertainty is only important for the lowest $\pt^{\mu\mu}$ range.

The global uncertainty of the integrated luminosity for the 2017 \pp data
is $1.6\%$, derived using methods described in Ref.~\cite{ATLAS-CONF-2019-021}.
Primary vertex association uncertainty, which results mostly from small discrepancies
between data and MC, was studied by varying the primary vertex association
requirements.
Since the primary-vertex association affects all Upsilon state yields in the
same way, it is treated as a global uncertainty together with
those of the \pp luminosity and $T_\mathrm{AA}$.
The combined systematic uncertainty for the luminosity and primary-vertex association
in \pp data is $2.6\%$.
For \PbPb collisions, the global systematic uncertainty of $\avgTAA$ is
estimated by
varying the Glauber model parameters as detailed in Ref.~\cite{HION-2016-06}.
The combined systematic uncertainty for $\avgTAA$ and primary-vertex association
in \PbPb collisions is $3.7\%$.
 
Systematic uncertainties in \pp and Pb+Pb collisions are summarized in Table~\ref{tab:totalsys}. While some systematic uncertainties for $R_{\mathrm{AA}}$ values and excited-state to ground-state double ratios are correlated (e.g.\ the acceptance) and cancel out in the ratios, most of the systematic uncertainties  are not completely correlated and are estimated by directly studying their effects on the ratios.
 
\begin{table}[bht]
\begin{center}
\caption{Summary of the sources of systematic uncertainty.}
\begin{tabular}{ c | c | c | c | c }
\hline
Collision type & Sources & \Upsi[\%] & \Upsins[\%] & \Upsins/\Upsi[\%] \\
\hline
\multirow{5}{*}{\pp collisions} & Luminosity & 1.6 & 1.6 & - \\
& Acceptance & 0.3--9.3 & 0.2--4.1 & - \\
& Efficiency & 2.7--7.0 & 2.8--4.0 & 3.0--7.1 \\
& Signal extraction & ~~3.1--10.2 & ~~4.3--11.9 & ~~4.5--12.2 \\
& Bin migration & <1~~~~~ & <1~~~~~ & - \\
& Primary-vertex association & 2.0 & 2.0 & - \\
 
\hline
\multirow{5}{*}{\PbPb collisions} & $\avgTAA$ & 0.8--8.2 & 0.8--8.2 & - \\
& Acceptance & 0.3--9.3 & 0.2--4.1 & - \\
& Efficiency & ~~4.0--15.0 & ~~3.9--25.3 & ~~4.4--28.8\\
& Signal extraction & ~~3.8--16.3 & 14.6--28.7 & 16.6--31.5 \\
& Bin migration & <2~~~~~ & <2~~~~~ & - \\
& Primary-vertex association & 3.4 & 3.4 & - \\
\hline
\end{tabular}
\label{tab:totalsys}
\end{center}
\end{table}
 
\section{Results}
 
\subsection{Differential cross-section}
 
Differential $\Upsins$ production cross-sections in \pp collisions are measured according to the relation
\begin{equation*}
\frac{d^2\sigma_{\Upsins}}{d\pt^{\mu\mu} dy^{\mu\mu}} \mathcal{B}(\Upsins\rightarrow\mu^+\mu^-) = \frac{N_{\Upsins}^\mathrm{corr}}{\mathrm{\Delta}\pt^{\mu\mu}\mathrm{\Delta} y^{\mu\mu} \int\!{\mathcal{L}dt}},
\end{equation*}
 
where $\mathcal{B}(\Upsins\rightarrow\mu^+\mu^-)$ is the dimuon decay
branching fraction, $N_{\Upsins}^\mathrm{corr}$ is the $\Upsins$ yield corrected
for acceptance and efficiencies,
$\Delta\pt^{\mu\mu}$ and $\Delta y^{\mu\mu}$ are the bin widths in $\pt^{\mu\mu}$ and $y^{\mu\mu}$,
and $\int\!{\mathcal{L}dt}$ is the integrated luminosity.
 
The \Upsins differential cross-sections in \pp collisions
at $5.02~\TeV$, multiplied by the respective dimuon branching fractions, are shown
as a function of $\pt^{\mu\mu}$ in the left panel of Figure~\ref{fig:cross}.

The per-event yields of \Upsins states in \PbPb collisions are defined by
 
\begin{equation*}
N_{\mathrm{AA}} = \frac{N_{\Upsins}^\mathrm{corr}}{\Delta\pt^{\mu\mu}\Delta y^{\mu\mu} N_{\mathrm{evt}}},
\end{equation*}
 
where $N_{\mathrm{evt}}$ is the total number of minimum-bias \PbPb collisions
in each centrality class.
In particular, this number is $1.02\times 10^{9}$ for the 0--10\% centrality interval.
Per-event Upsilon yields in \PbPb collisions divided by $\avgTAA$
are shown in the right panel of Figure~\ref{fig:cross}.
The results for \Ups{3}
mesons are not shown because their peaks are not statistically significant in \PbPb collisions.
 
\begin{figure}[htb]
\begin{center}
\includegraphics[width=\fourfigsize]{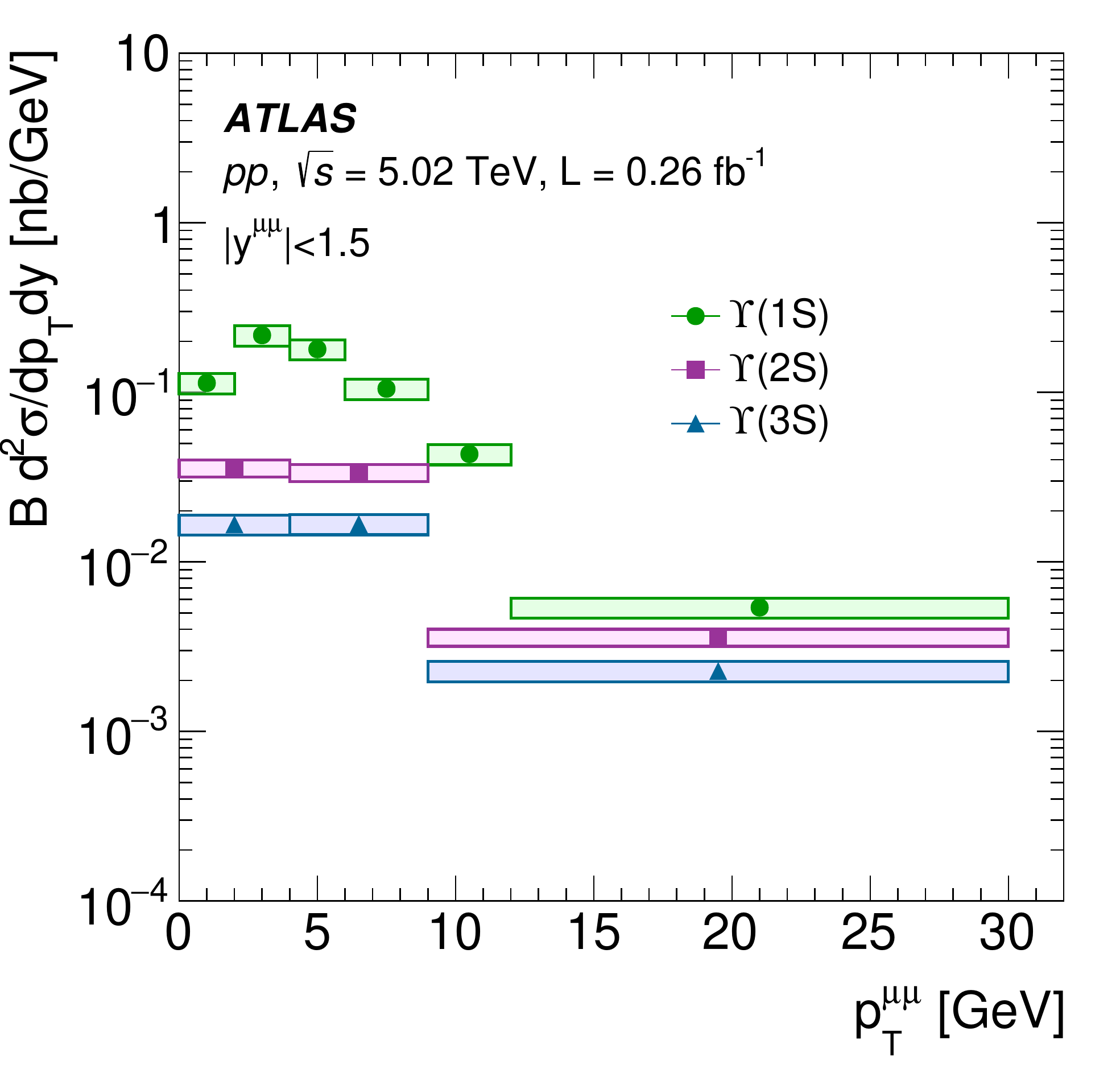}
\includegraphics[width=\fourfigsize]{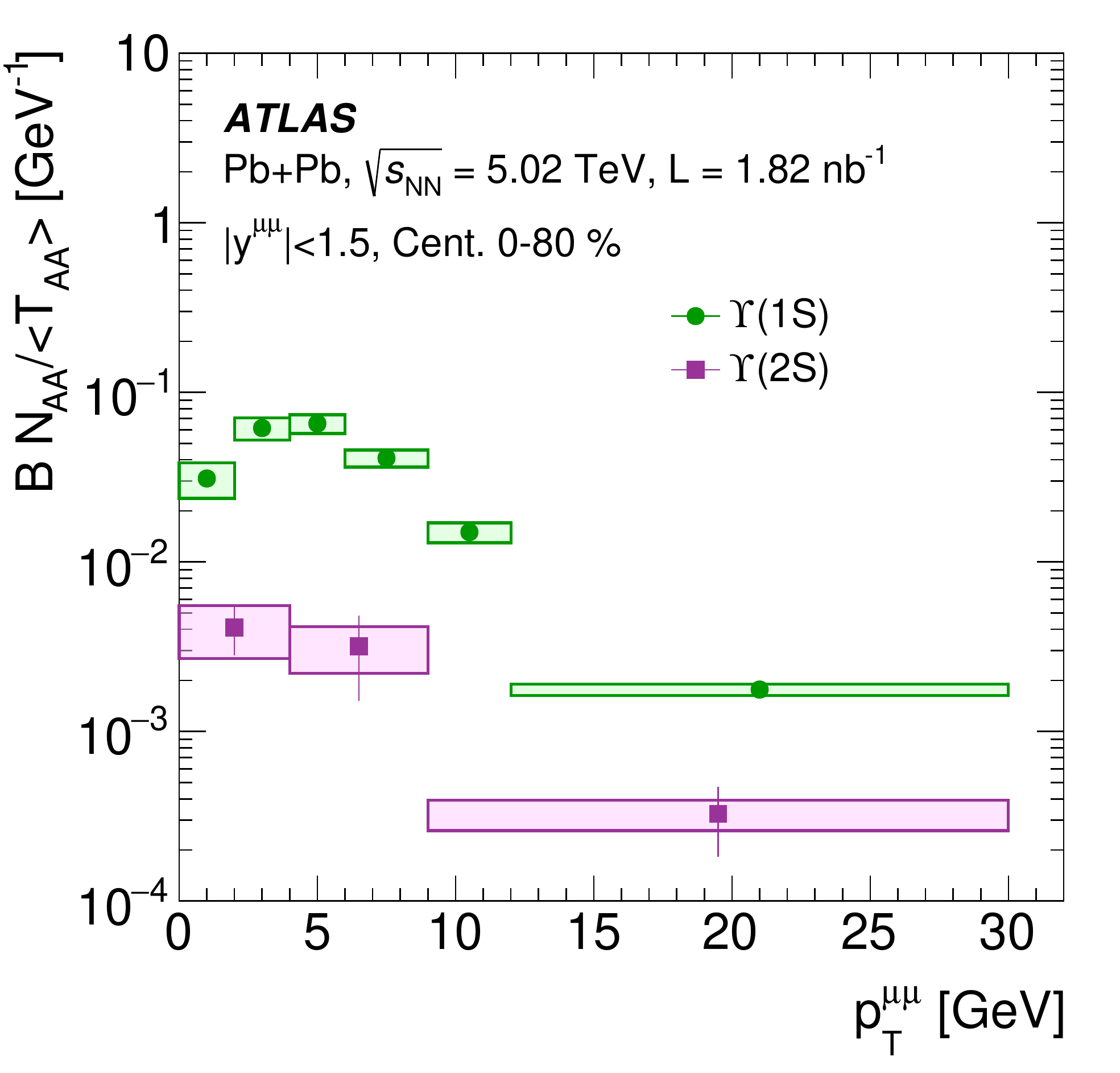}
\caption{Production cross-sections of \Ups{1}, \Ups{2}, and \Ups{3} mesons as a function of $\pt^{\mu\mu}$ in \pp collisions (left) and per-event yields in \PbPb collisions (right) at 5.02~\TeV. Error bars indicate the statistical uncertainties and boxes represent the systematic uncertainties. Not shown are the correlated systematic uncertainties of $2.6\%$ for luminosity and primary-vertex association in \pp and $3.7\%$ for $\avgTAA$ and primary-vertex association in \PbPb collisions.}
\label{fig:cross}
\end{center}
\end{figure}
 
\subsection{Nuclear modification factor}
 
The modifications of bottomonium production yields in \PbPb collisions relative
to the \pp system are quantified by the nuclear modification factor \RAA, which can
be defined for each centrality interval as
 
\begin{equation*}
\RAA = \frac{N_\mathrm{AA}}{\avgTAA \sigma^{\pp}},
\end{equation*}
where $N_\mathrm{AA}$ is the observed per-event yield of bottomonium states in \PbPb collisions,
and $\sigma^{\pp}$
is the bottomonium production cross-section in \pp collisions at the same collision energy.
 
Figure~\ref{fig:raa} shows the \RAA values of $\Upsins$ as functions of \avgNpart
(top), dimuon $p_\mathrm{T}^{\mu\mu}$ (bottom left), and $|y^{\mu\mu}|$ (bottom right).
The centrality-integrated results are also shown in the right panel of the top plot.
In addition to the results for $\Upsi$ and $\Upsii$, only the combined result for
the two excited states, $\varUpsilon(\textrm{2S+3S})$, is
presented because the $\Upsiii$ peak is not statistically significant in the \PbPb data.
The $\Upsins$ states are observed to be suppressed over the whole kinematic range
investigated, and the \RAA values of \Upsii and \Upsiex are always lower than those
of \Upsi. The \RAA value decreases with \avgNpart for all three states.
No strong $\pt^{\mu\mu}$ or $|y^{\mu\mu}|$ dependence is observed. When no statistically significant non-zero
yield was extracted for a particular kinematic selection, 95\% confidence
level upper limit was calculated.
 
\begin{figure}[htb]
\begin{center}
\includegraphics[width=\fourfigsize]{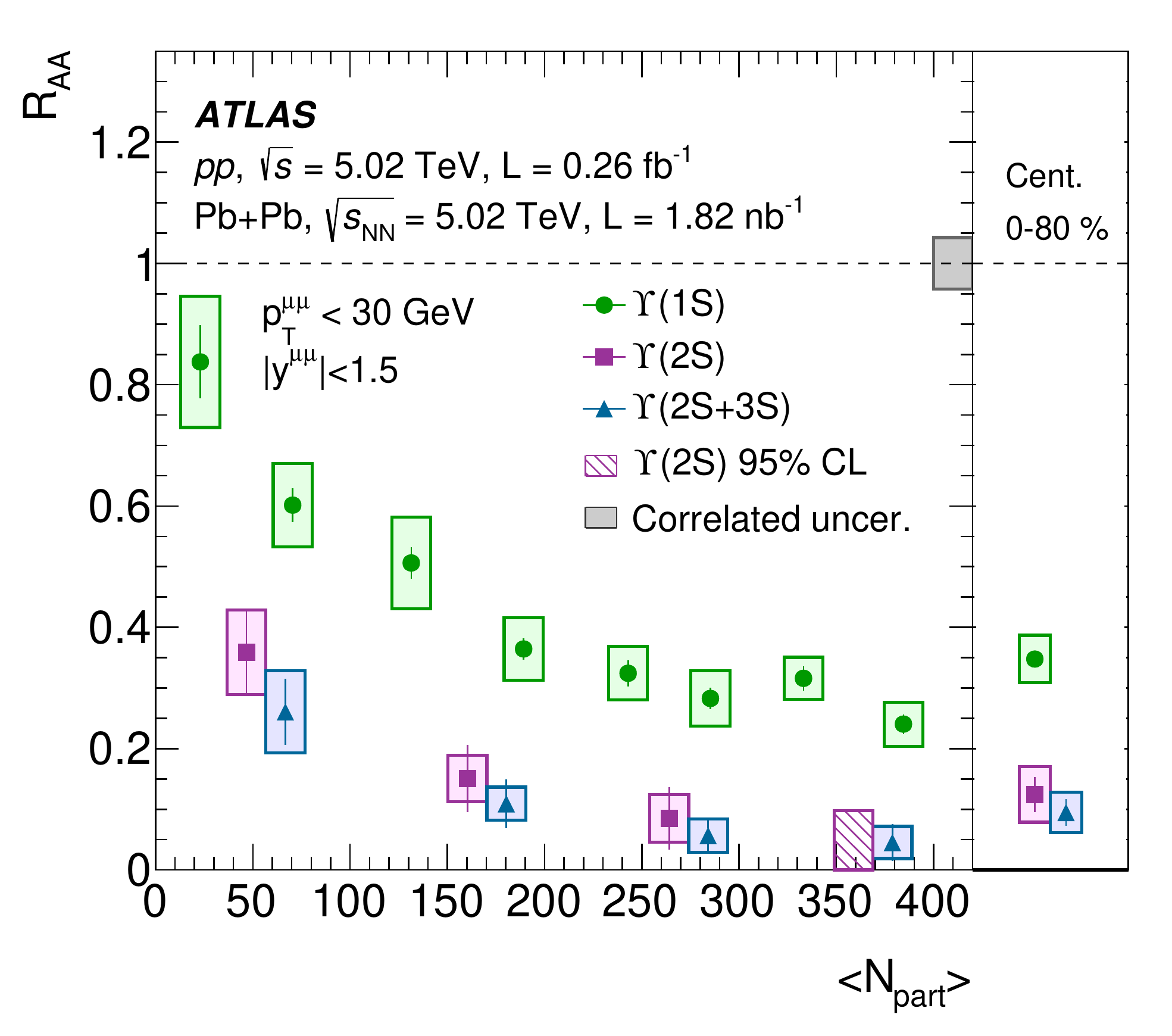}\\
\includegraphics[width=\fourfigsize]{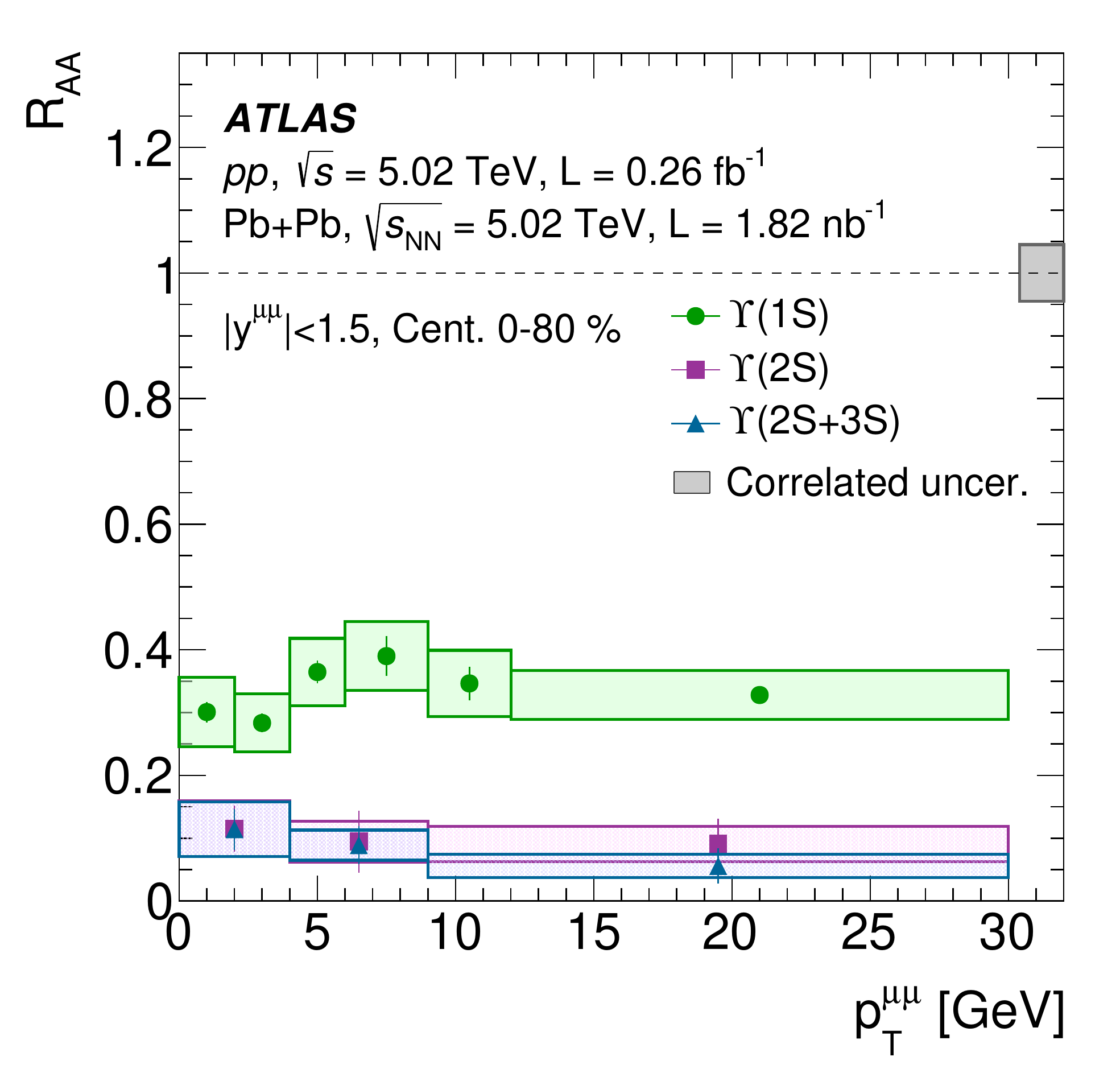}
\includegraphics[width=\fourfigsize]{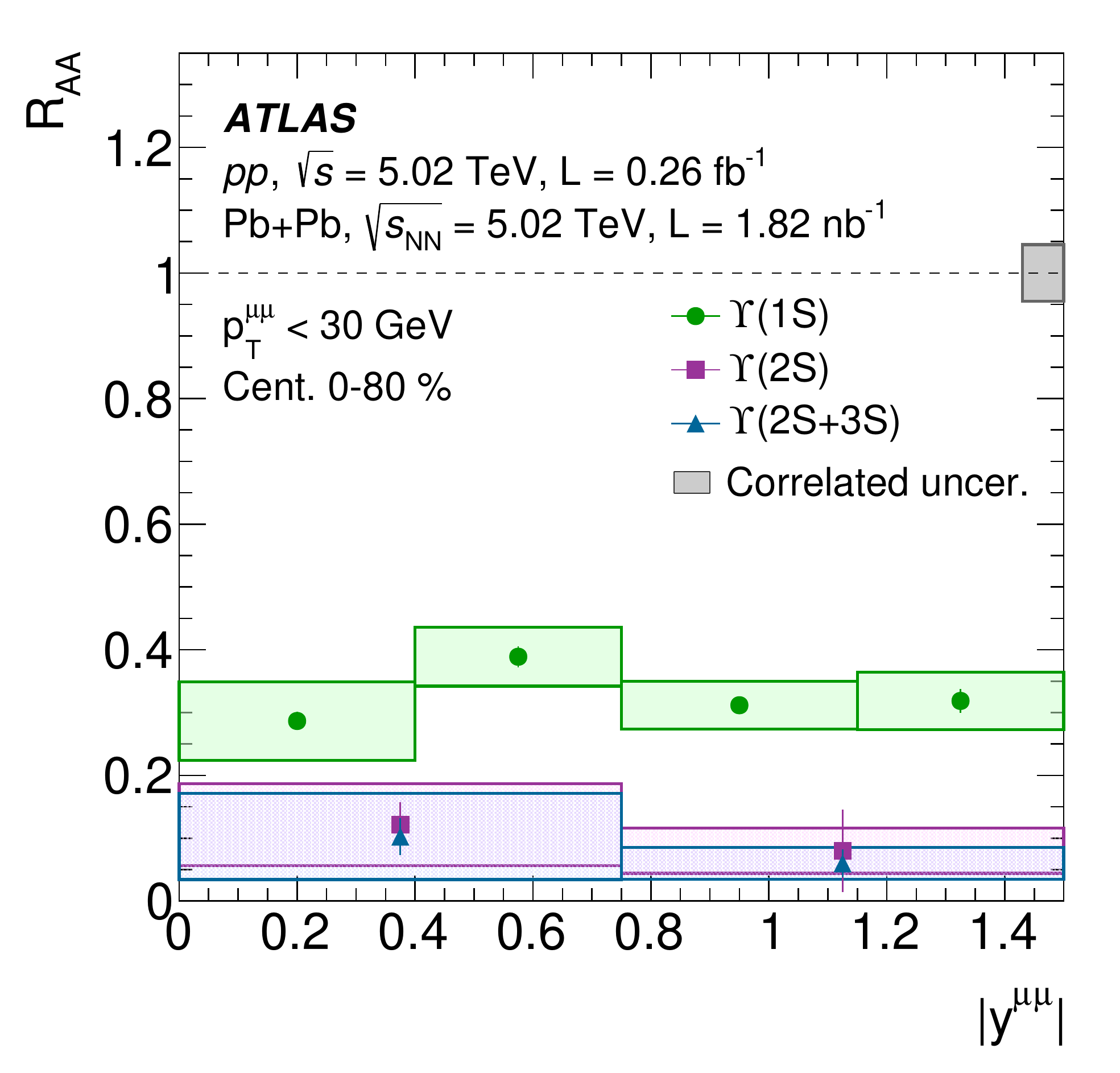}
\caption{The nuclear modification factor \RAA of $\Upsi$, $\Upsii$, and $\varUpsilon(\textrm{2S+3S})$ as functions of centrality (top),  $\pt^{\mu\mu}$ (bottom left), and $|y^{\mu\mu}|$ (bottom right) at $5.02~\TeV$. Error bars indicate the statistical uncertainties and boxes represent the systematic uncertainties. The gray boxes around $\RAA=1$ correspond to the global systematic uncertainty. The right panel of the top plot shows the \RAA results integrated over centrality.}
\label{fig:raa}
\end{center}
\end{figure}
 
\subsection{Excited-state to ground-state double ratios}
 
The suppression of different Upsilon states can be compared by constructing
an excited-state to ground-state double ratio of nuclear modification factors.
The advantage of measuring the double ratios is that
the acceptance and efficiency corrections partially cancel out, and the overall
systematic
uncertainty is reduced.
Although defined in terms of the individual nuclear suppression factors, the double ratio can be understood as being defined as the ratio
of the yields of excited states \Upsii, \Upsiii
or of combined yield of the two excited states ($\varUpsilon(2S+3S))$
to the yield of the ground state \Upsi in \PbPb
collisions, divided by the same ratio in \pp collisions.
\begin{equation*}
\rho_{\mathrm{AA}}^{\Upsins/\Upsi} = \RAA(\Upsins)/\RAA(\Upsi).
\end{equation*}
 
Figure~\ref{fig:dbl} shows the
\rhoaan for \Upsii and \Upsiex as
functions of $N_{\textrm{part}}$ (top), $\pt^{\mu\mu}$ (bottom left),
and $|y^{\mu\mu}|$ (bottom right).
The centrality-integrated results are also shown in the right panel of the top plot.
The $\rho_{\mathrm{AA}}^{\Upsins/\Upsi}$ values for \Upsii and \Upsiex are always less
than one, indicating the excited states are more suppressed than the ground state.
The centrality-dependent $\rho_{\mathrm{AA}}^{\Upsins/\Upsi}$ shows a slightly decreasing trend toward more central collisions, but no $\pt^{\mu\mu}$ or $|y^{\mu\mu}|$ dependence is observed.
 
These results are consistent with previous measurements of Upsilon
suppression at LHC energies by the CMS~\cite{CMS-HIN-16-023}
and ALICE~\cite{ALICE-Ups-2014} experiments.
 
\begin{figure}[htb]
\begin{center}
\includegraphics[width=\fourfigsize]{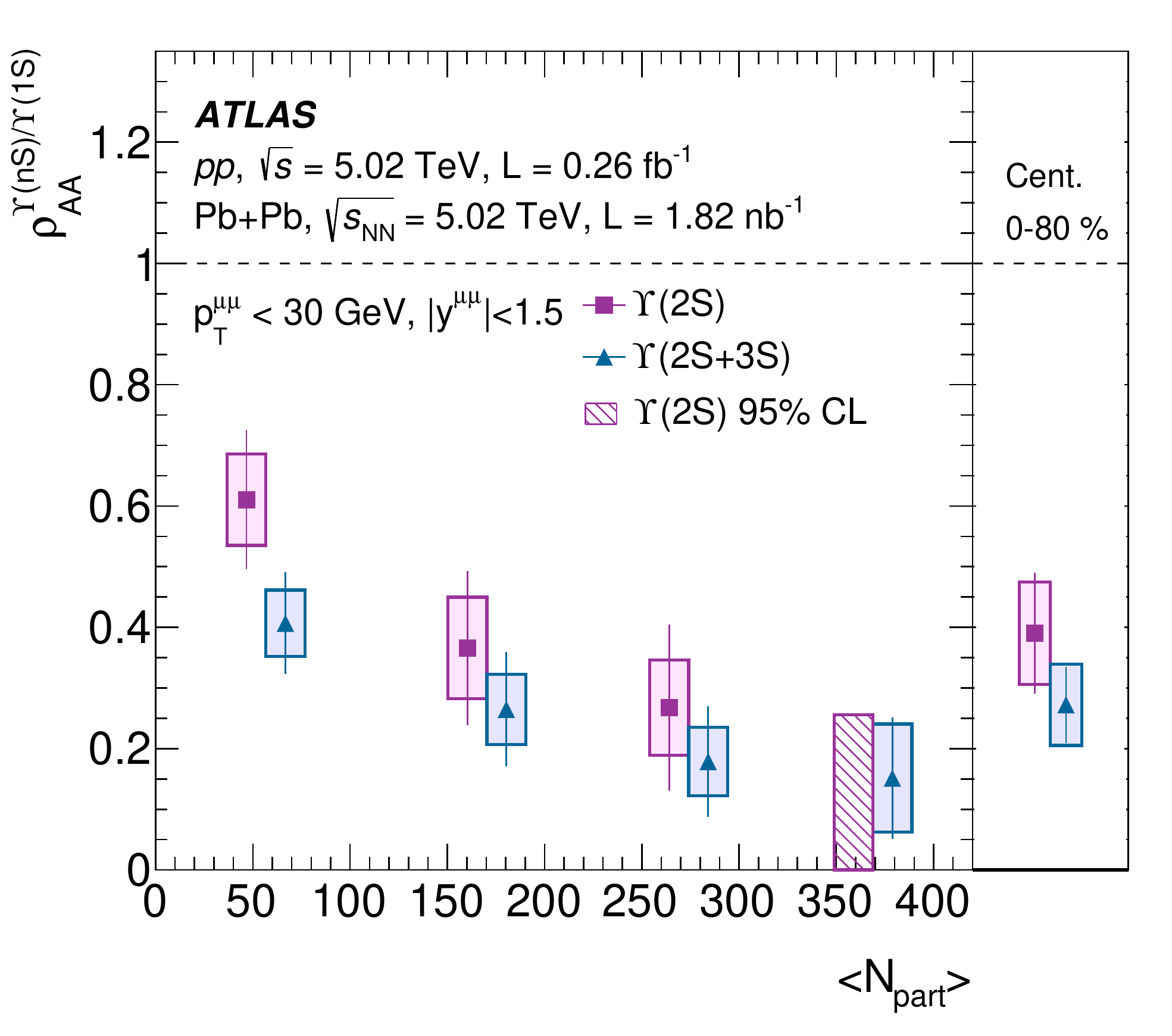}\\
\includegraphics[width=\fourfigsize]{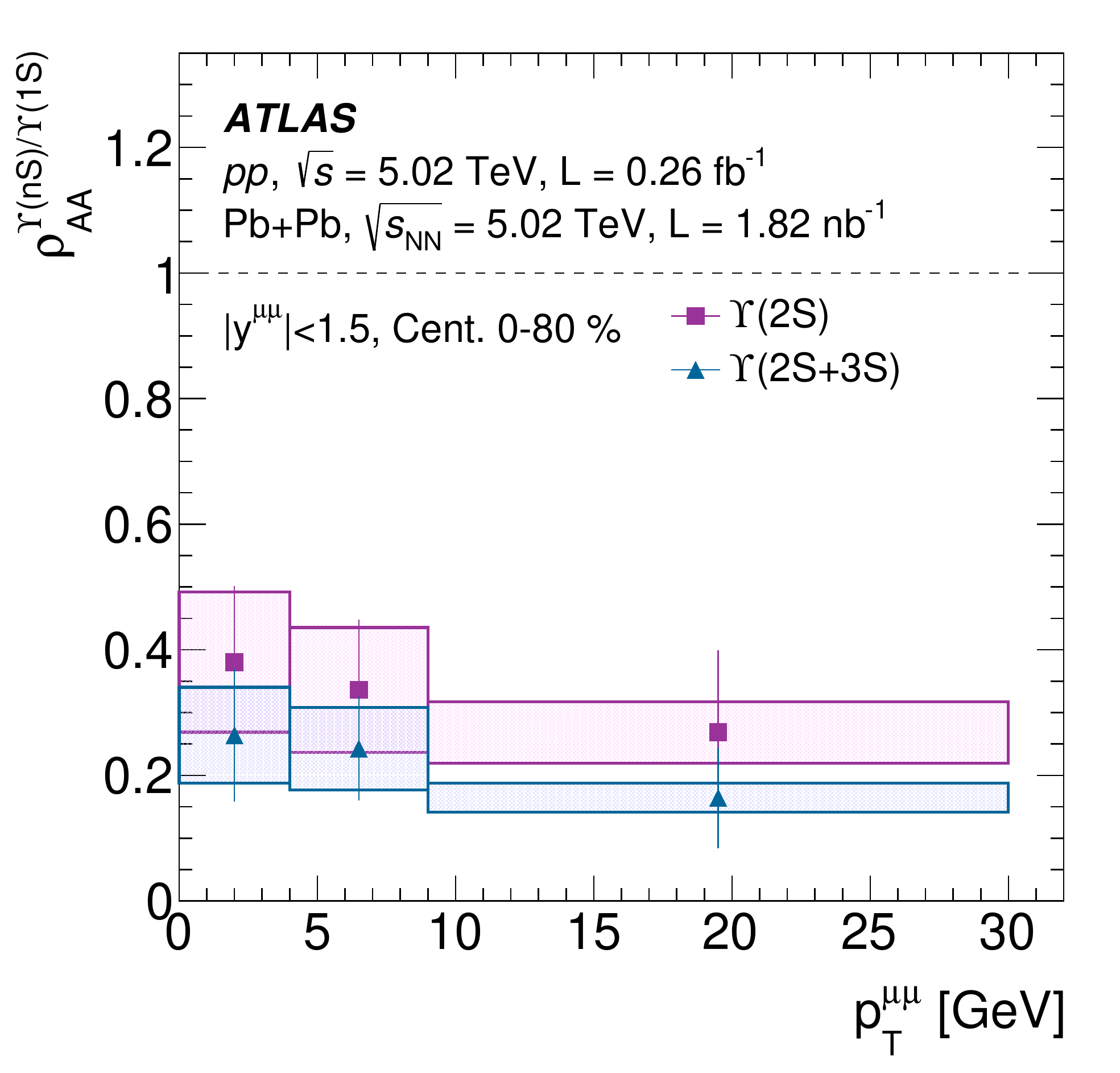}
\includegraphics[width=\fourfigsize]{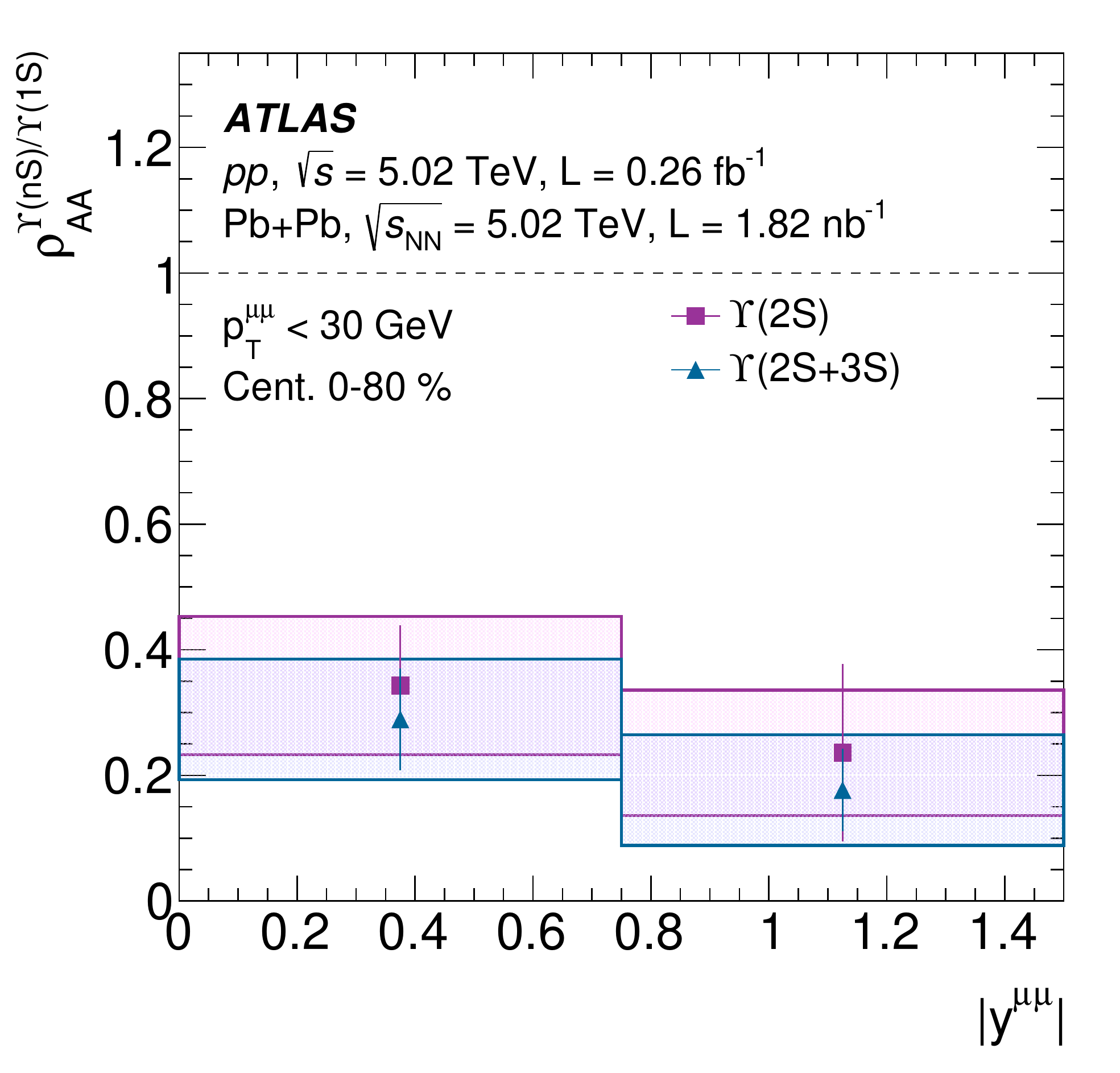}
\caption{The double ratio $\rho_{\mathrm{AA}}^{\Upsins/\Upsi}$ for $\Upsii$ and $\varUpsilon(\textrm{2S+3S})$ as functions of centrality (top), $\pt^{\mu\mu}$ (bottom left), and $|y^{\mu\mu}|$ (bottom right) at $5.02~\TeV$ per nucleon--nucleon pair. Error bars indicate the statistical uncertainties and boxes represent the systematic uncertainties. The right panel of the top plot shows the results integrated over centrality.}
\label{fig:dbl}
\end{center}
\end{figure}
 
\subsection{Theory comparisons and discussion}
 
Figure~\ref{fig:strickland_raa} shows the \RAA of \Upsi and \Upsii and the
double ratio \rhoaatwo compared with a calculation by N.~Brambilla
et al.\ in Ref.~\cite{Strickland:2021}. This model uses  potential NRQCD and the
formalism of open quantum systems to numerically solve the Lindblad equation using a stochastic unraveling called the quantum trajectories algorithm.
Heavy-quark interactions with the strongly coupled medium are encoded in the
two nonperturbative transport coefficients: the heavy-quark momentum diffusion
coefficient and its dispersive counterpart. The authors of Ref.~\cite{Strickland:2021} have run variations of these parameters within reasonable ranges.
 
\begin{figure}[htb]
\begin{center}
\includegraphics[width=\fourfigsize]{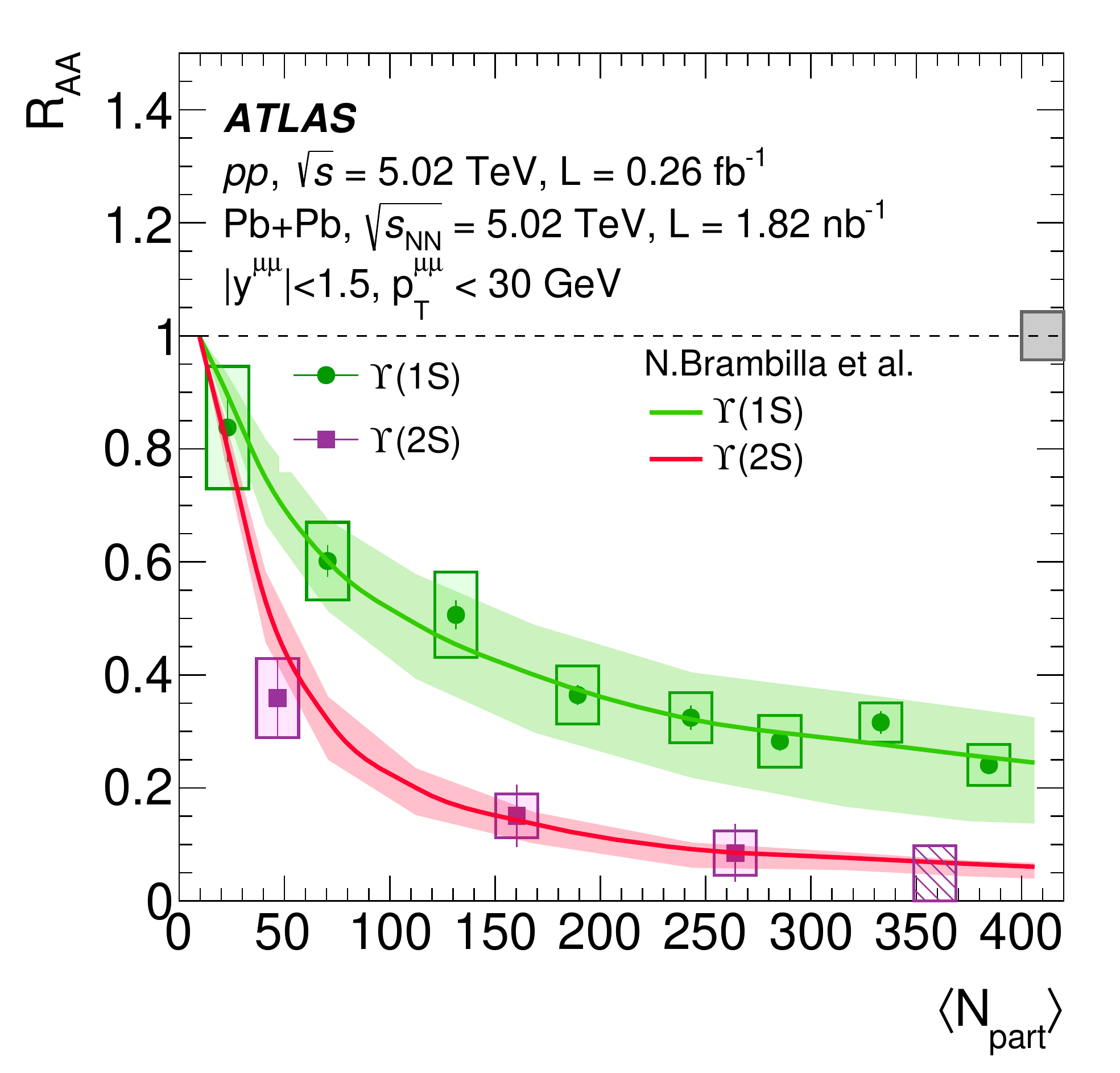}
\includegraphics[width=\fourfigsize]{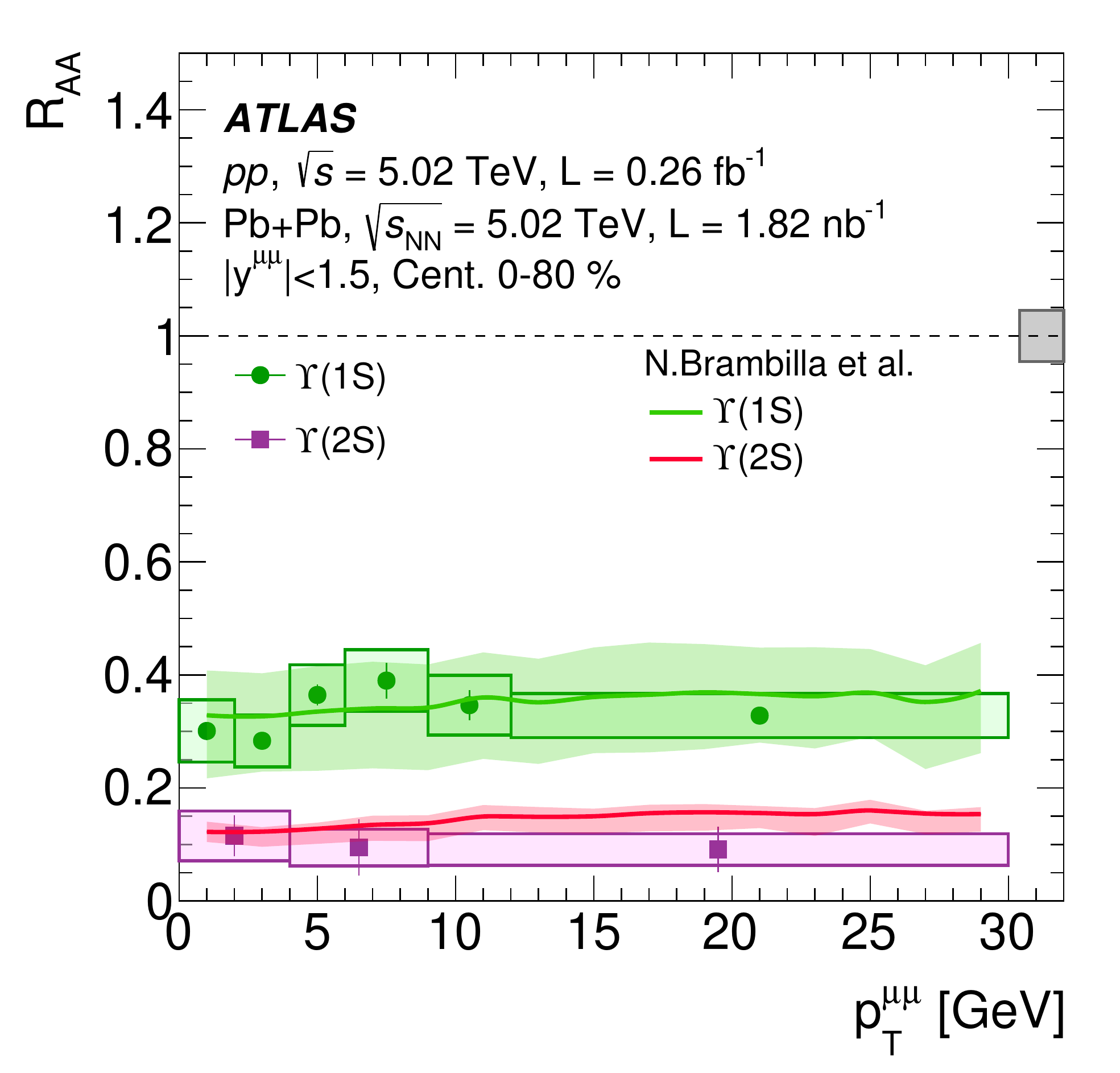}
\includegraphics[width=\fourfigsize]{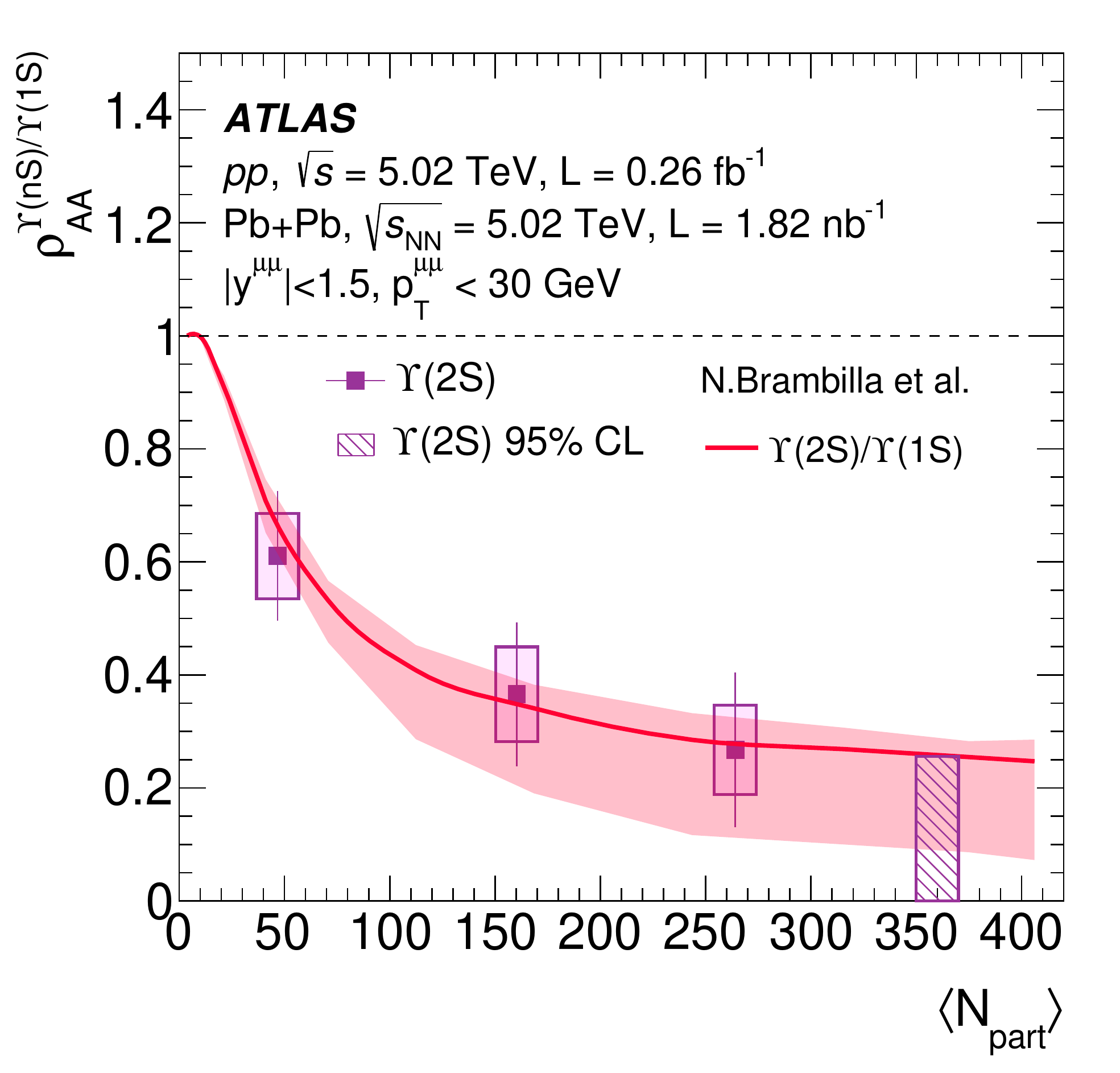}
\includegraphics[width=\fourfigsize]{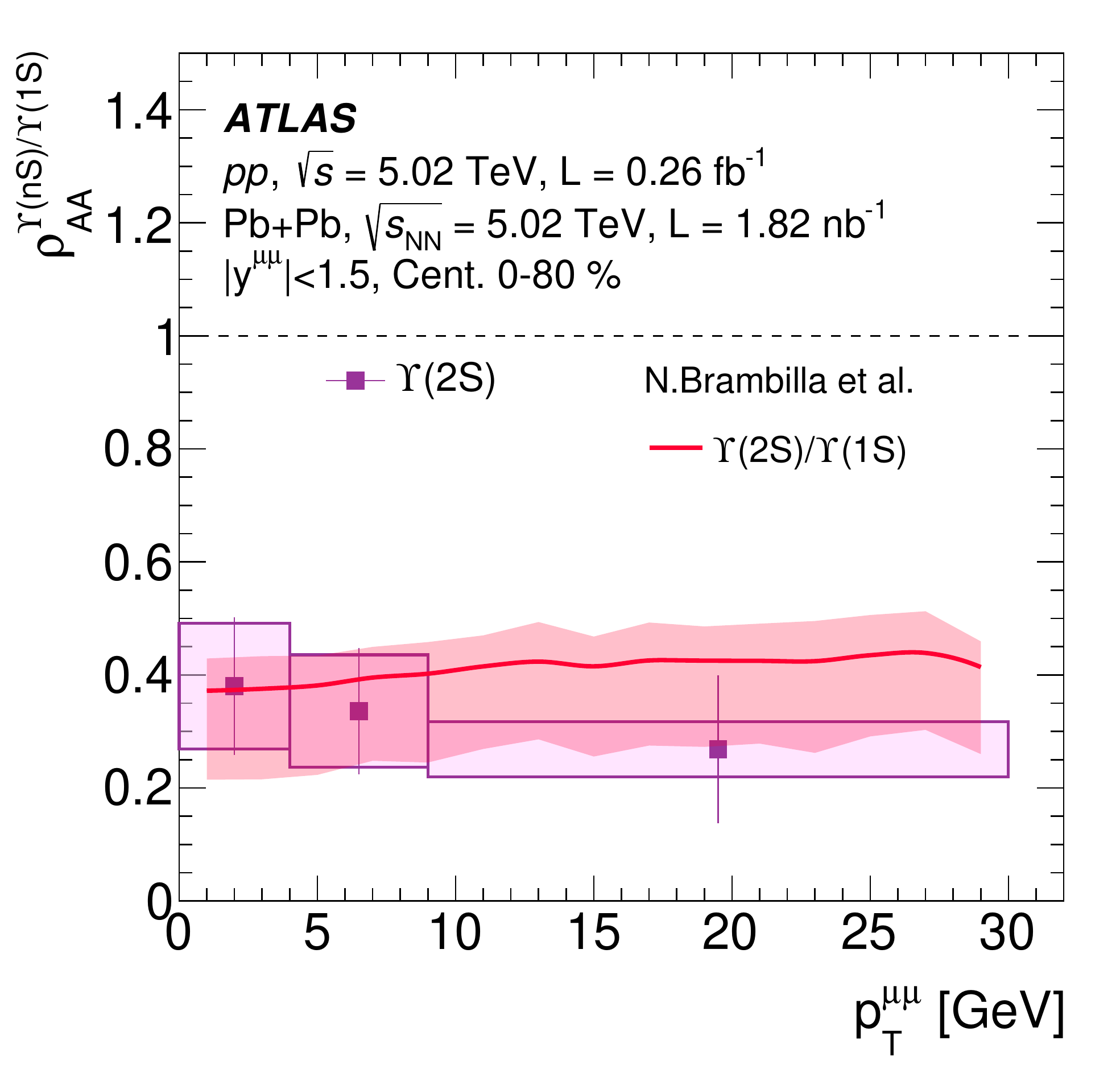}
\caption{The nuclear modification factor \RAA of $\Upsi$ and  $\Upsii$ (top row) and the double ratio $\rho_{\mathrm{AA}}^{\Upsins/\Upsi}$ for $\Upsi$ and $\Upsii$ (bottom row) as functions of centrality (left column) and  $\pt^{\mu\mu}$ (right column) at $5.02~\TeV$ per nucleon--nucleon pair compared to a calculation by N.~Brambilla et al., Ref.~\cite{Strickland:2021} (solid curves). Color bands represent model uncertainties due to variation of the model parameters.}
\label{fig:strickland_raa}
\end{center}
\end{figure}

Figure~\ref{fig:du} shows the \RAA of \Upsi, \Upsii and \Upsiex,
and the double ratio \rhoaan for \Upsii and \Upsiex
compared to a calculation by X.~Du et al.\ in Ref.~\cite{Du:2017}.
This model uses a kinetic-rate equation approach including regeneration and
has four dimensionless parameters which characterize the
temperature dependence of the pertinent screening masses.
These parameters are extracted through the fits to the data already available
at RHIC and LHC.
The
band corresponding to the 95\% confidence interval
is shown in the figure.
 
\begin{figure}[htb]
\begin{center}
\includegraphics[width=\fourfigsize]{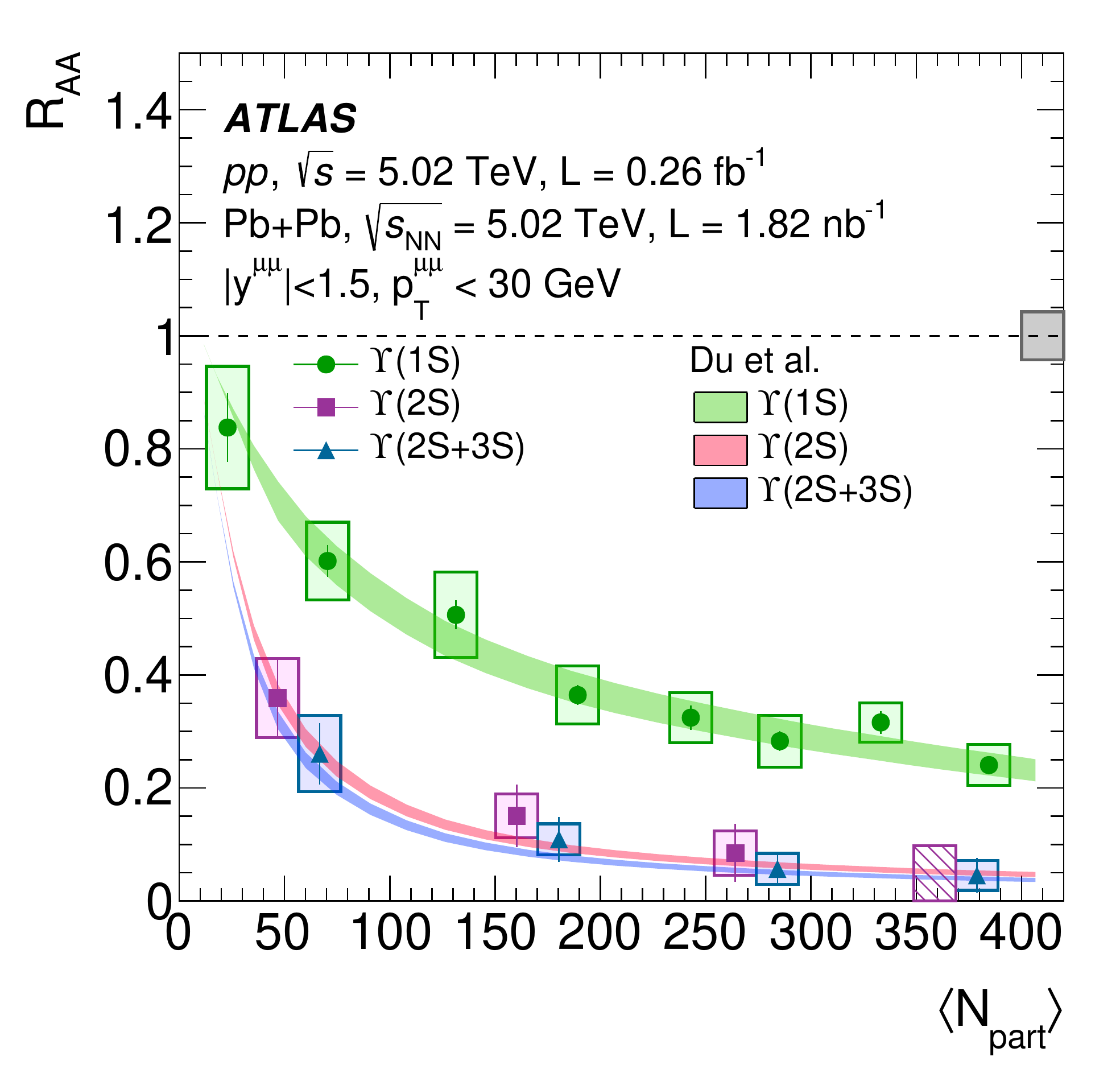}
\includegraphics[width=\fourfigsize]{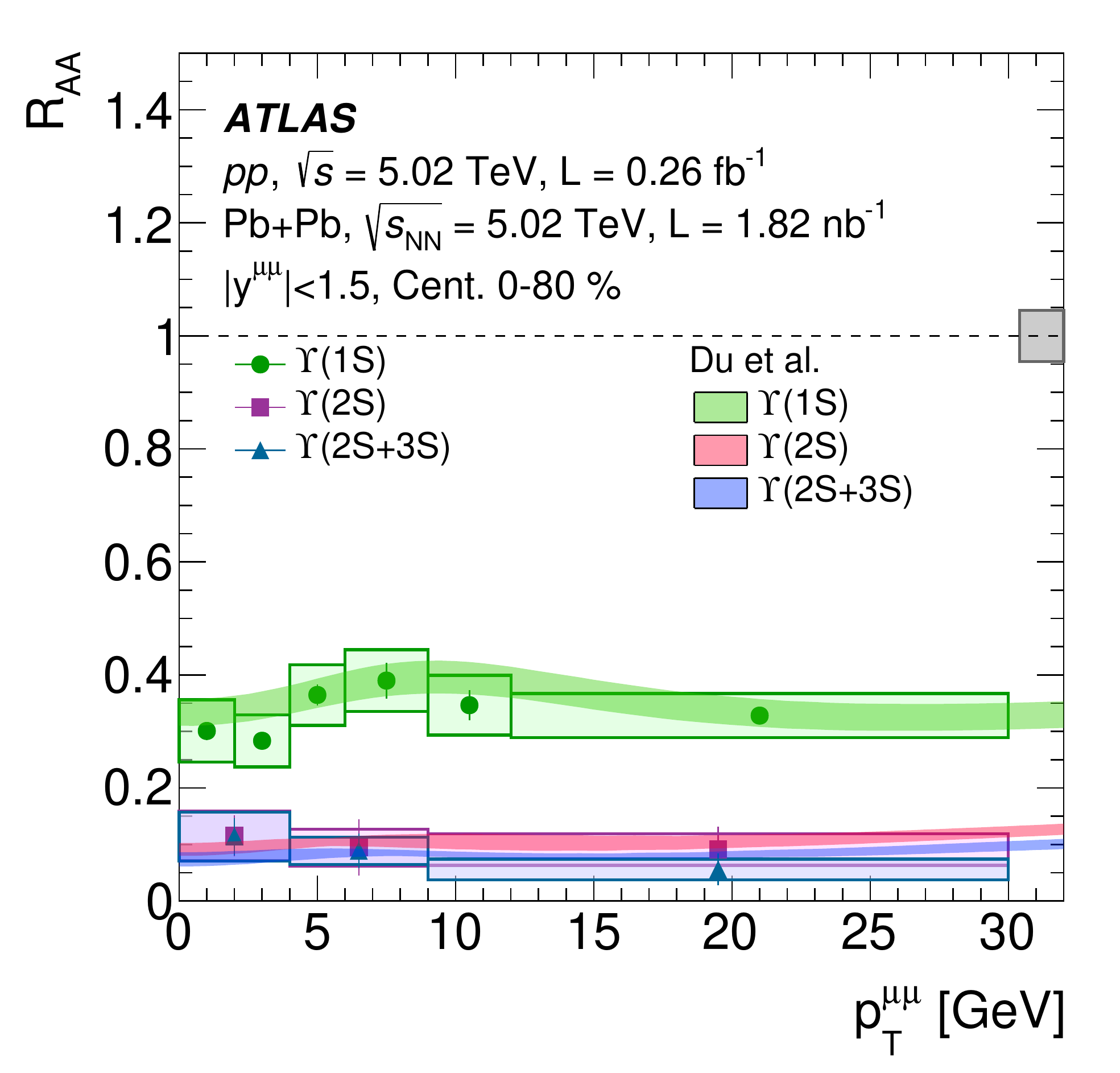}
\includegraphics[width=\fourfigsize]{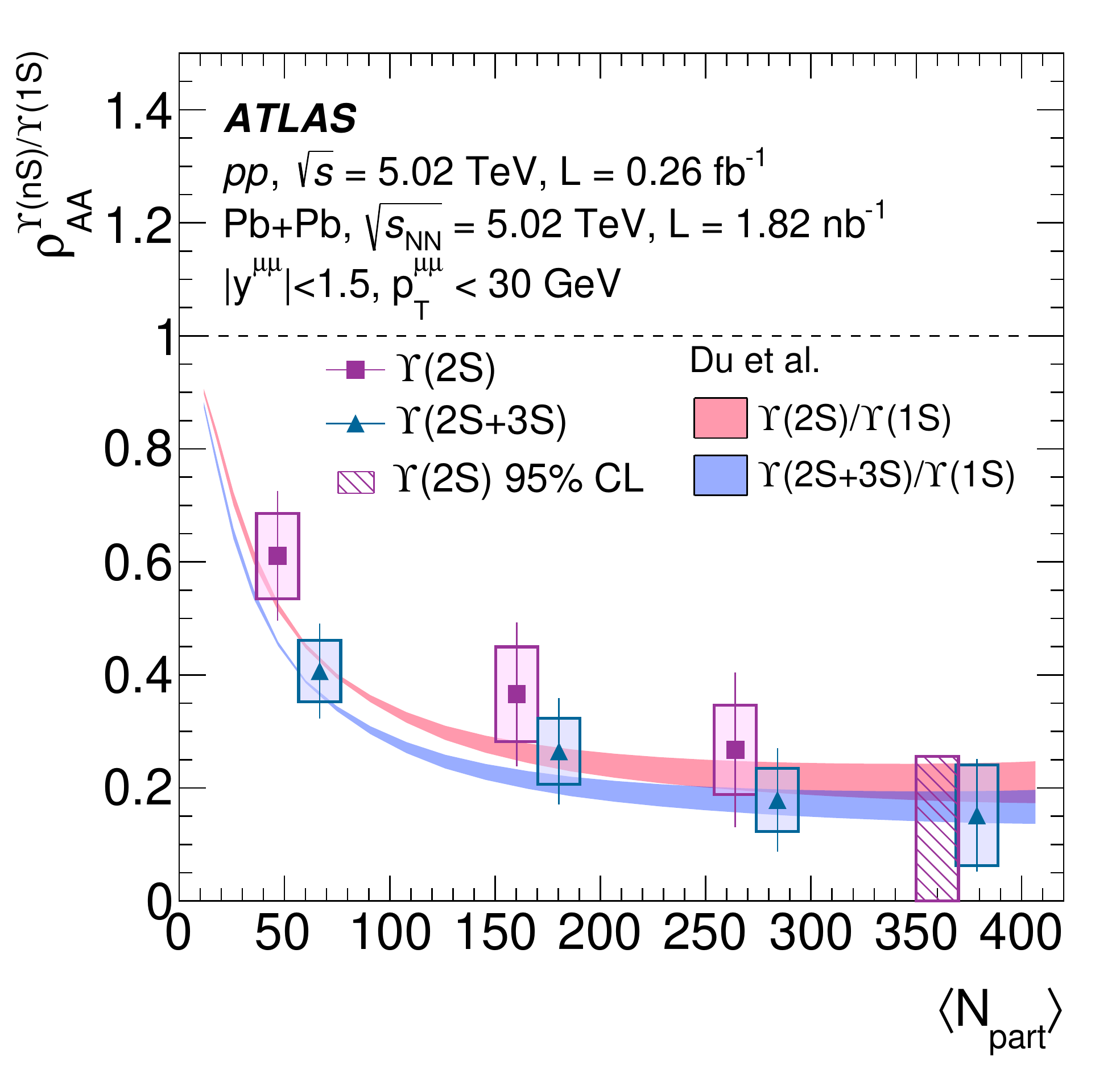}
\includegraphics[width=\fourfigsize]{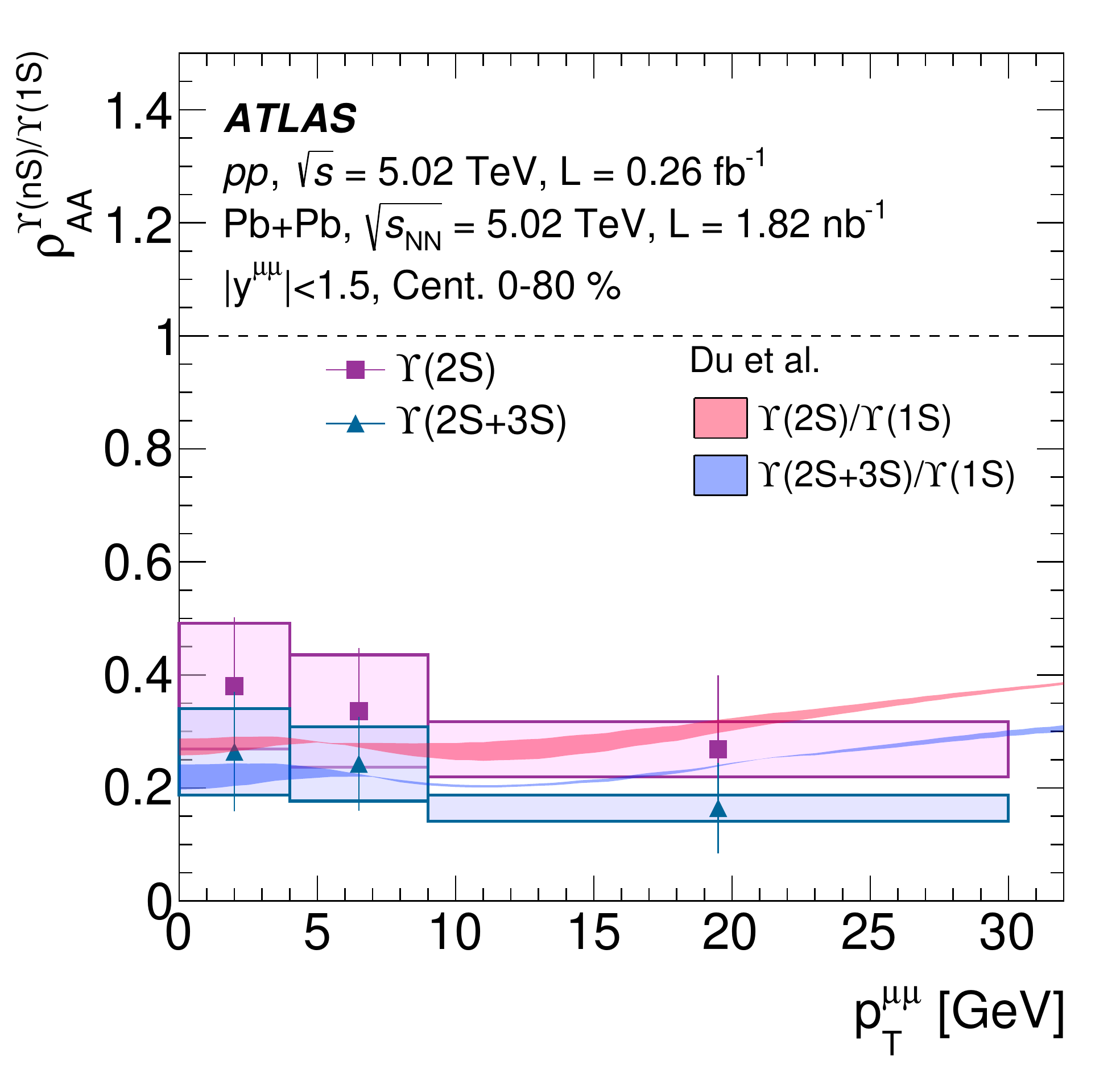}
\caption{The nuclear modification factor \RAA of $\Upsi$ and  $\Upsii$ (top row) and the double ratio $\rho_{\mathrm{AA}}^{\Upsins/\Upsi}$ for $\Upsii$ (bottom row) as functions of centrality (left) and $\pt^{\mu\mu}$ (right) at $5.02~\TeV$ per nucleon--nucleon pair compared to a calculation by X.~Du et al., Ref.~\cite{Du:2017}. The bands represent 95\% confidence level (CL) limits.}
\label{fig:du}
\end{center}
\end{figure}
 
Figure~\ref{fig:yao_raa} shows the \RAA of \Upsi, \Upsii and \Upsiex,
as well as the double ratio $\rho_{\mathrm{AA}}^{\Upsins/\Upsi}$
for $\Upsi$ and $\Upsii$,
compared to a calculation by X.~Yao et al.\ in Ref.~\cite{Yao:2021}.
This model uses a framework with coupled transport equations for
open heavy-flavor
and quarkonium states in order to describe their transport
inside the quark--gluon plasma, including regeneration.
Cold nuclear matter effects are included by using nuclear parton distribution
functions for the initial primordial heavy-flavor production.
A calibrated $(2 + 1)$-dimensional viscous hydrodynamic model is used to describe
the bulk QCD medium. The model depends on the choice of nucleus parton
distribution function (nPDF) and two coupling constant parameters, $\alpha_\text{s}$ and $\alpha_\text{s}^\text{pot}$, which are varied by $\pm10$\% from their nominal values.
 
\begin{figure}[htb]
\begin{center}
\includegraphics[width=0.43\columnwidth]{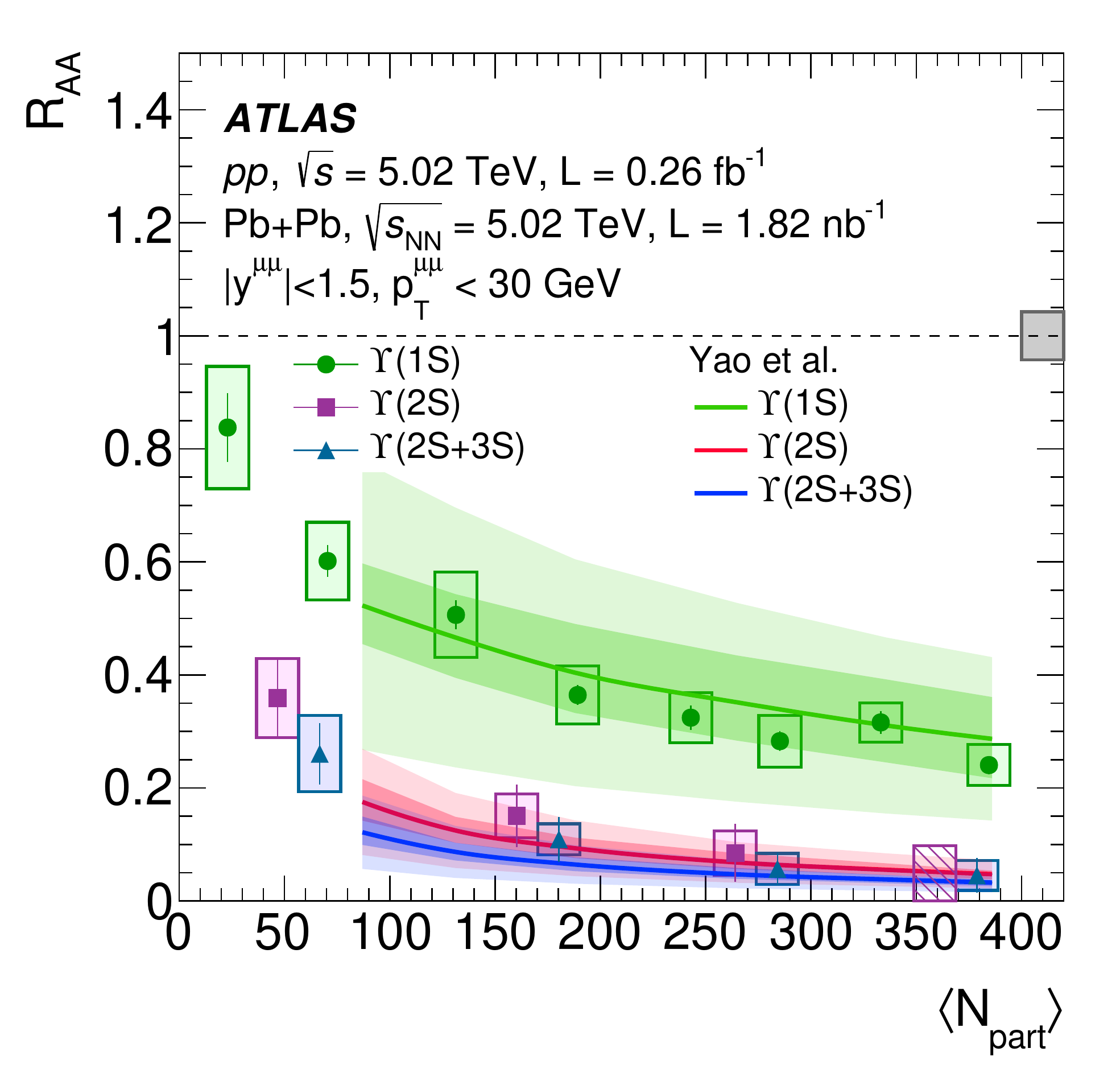}
\includegraphics[width=0.43\columnwidth]{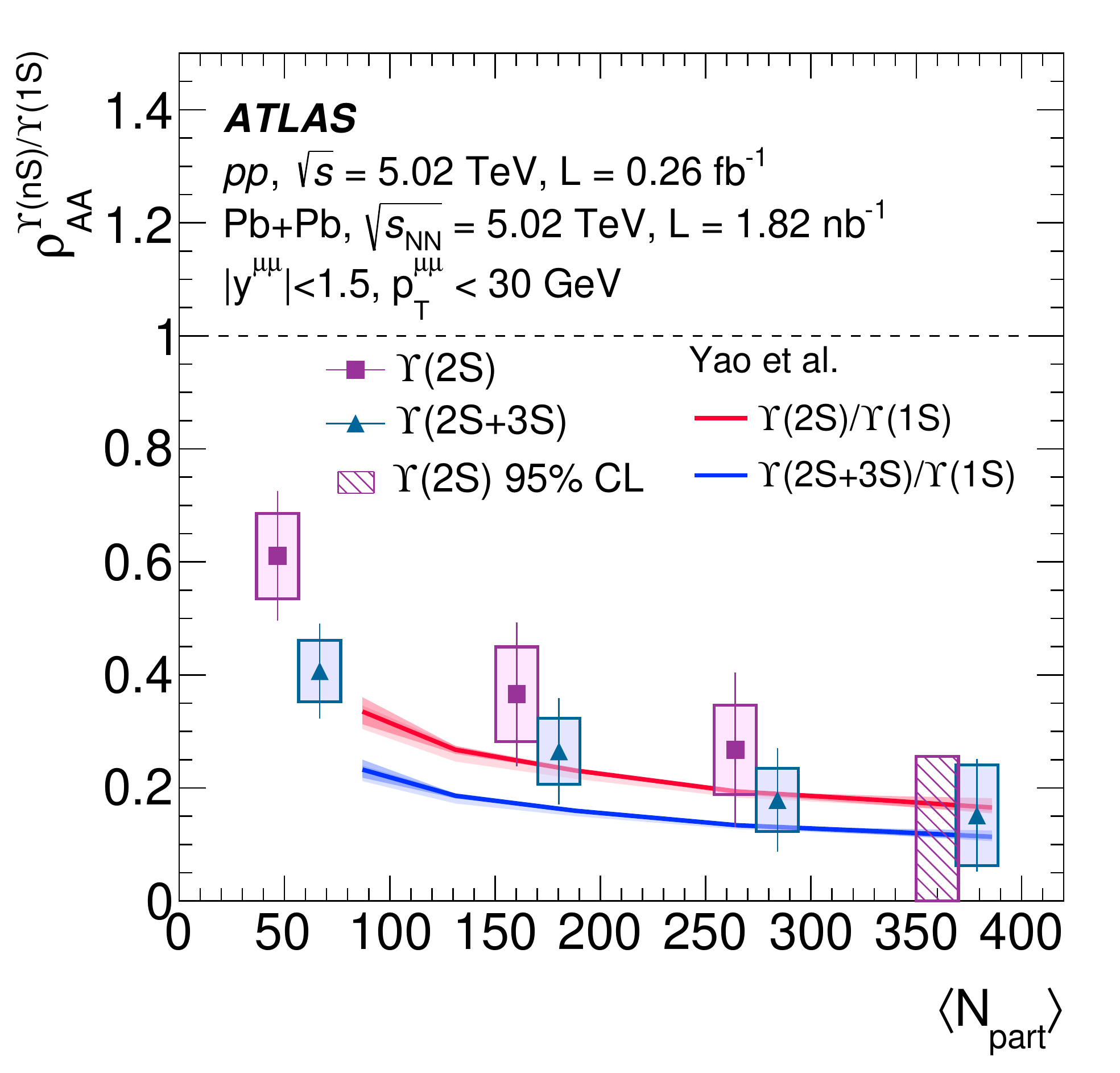}
\includegraphics[width=0.43\columnwidth]{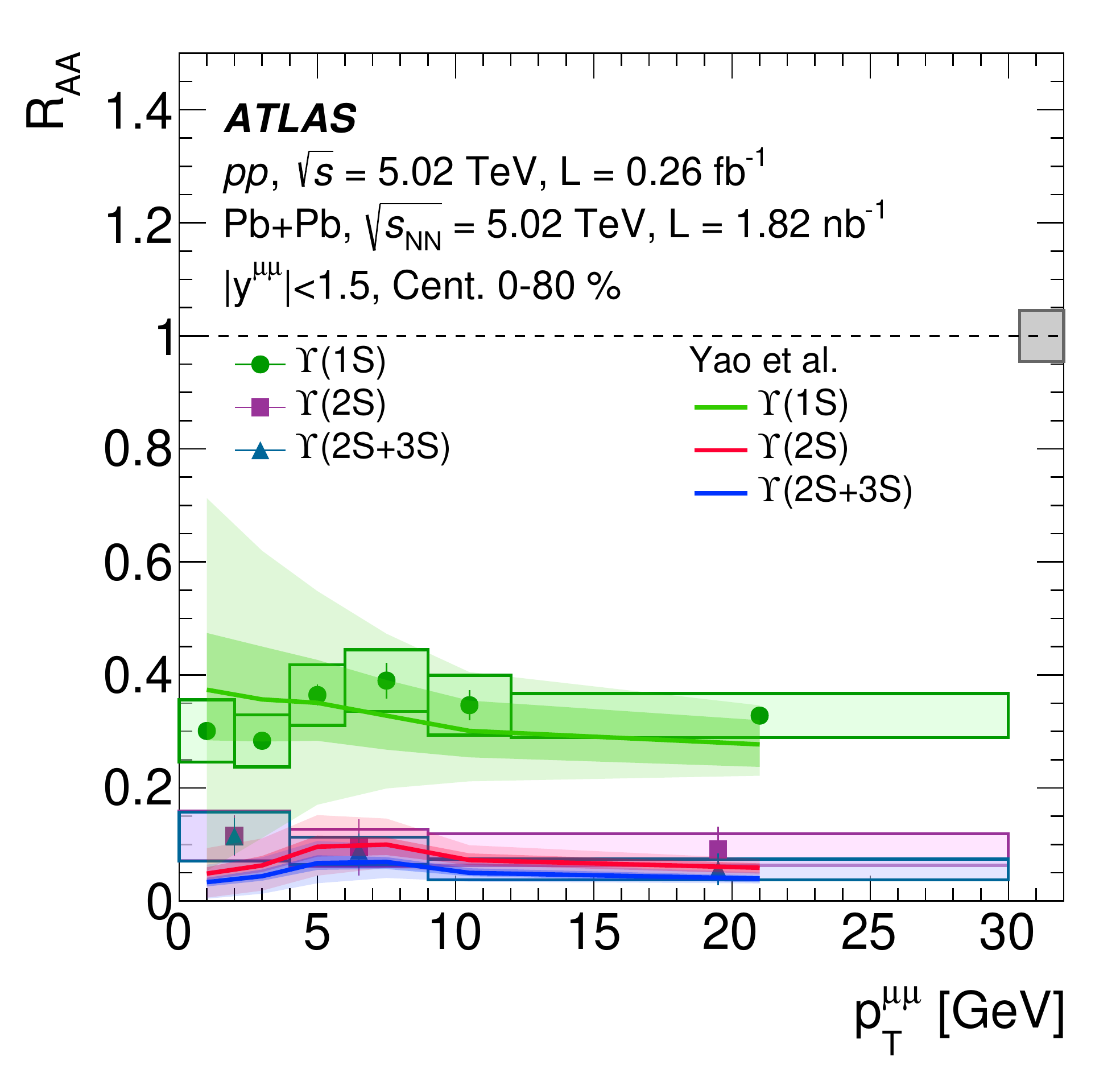}
\includegraphics[width=0.43\columnwidth]{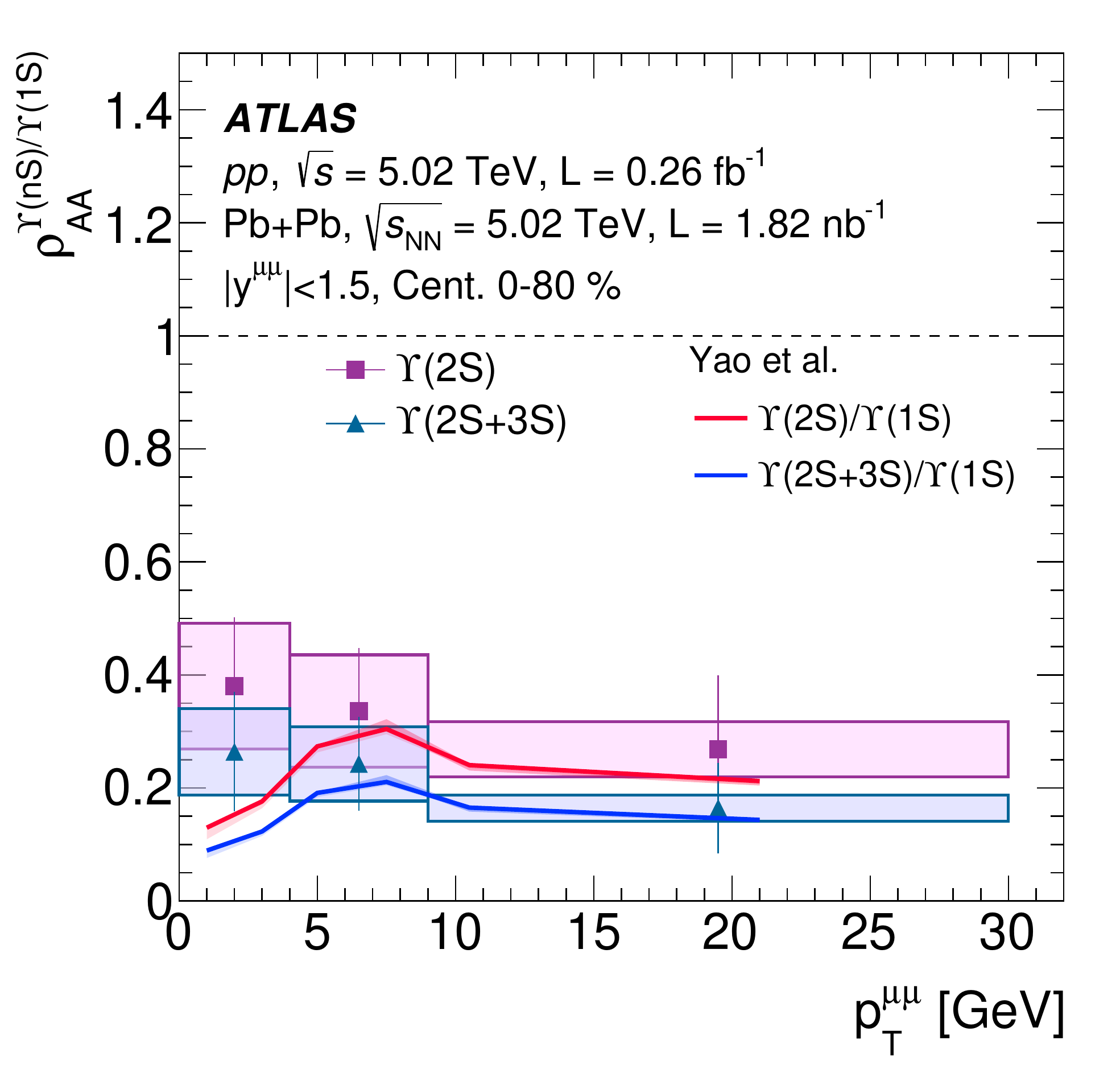}
\includegraphics[width=0.43\columnwidth]{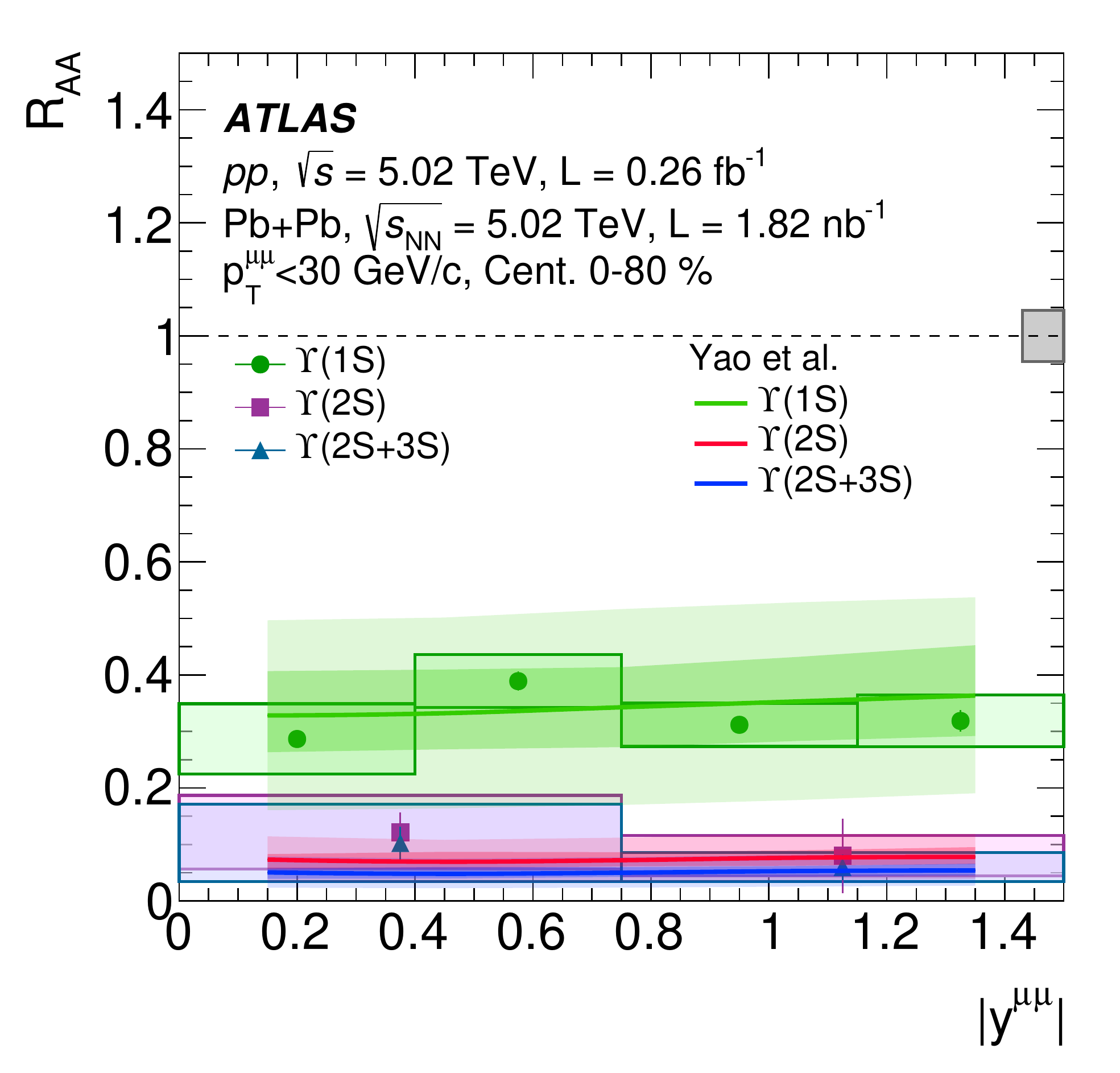}
\includegraphics[width=0.43\columnwidth]{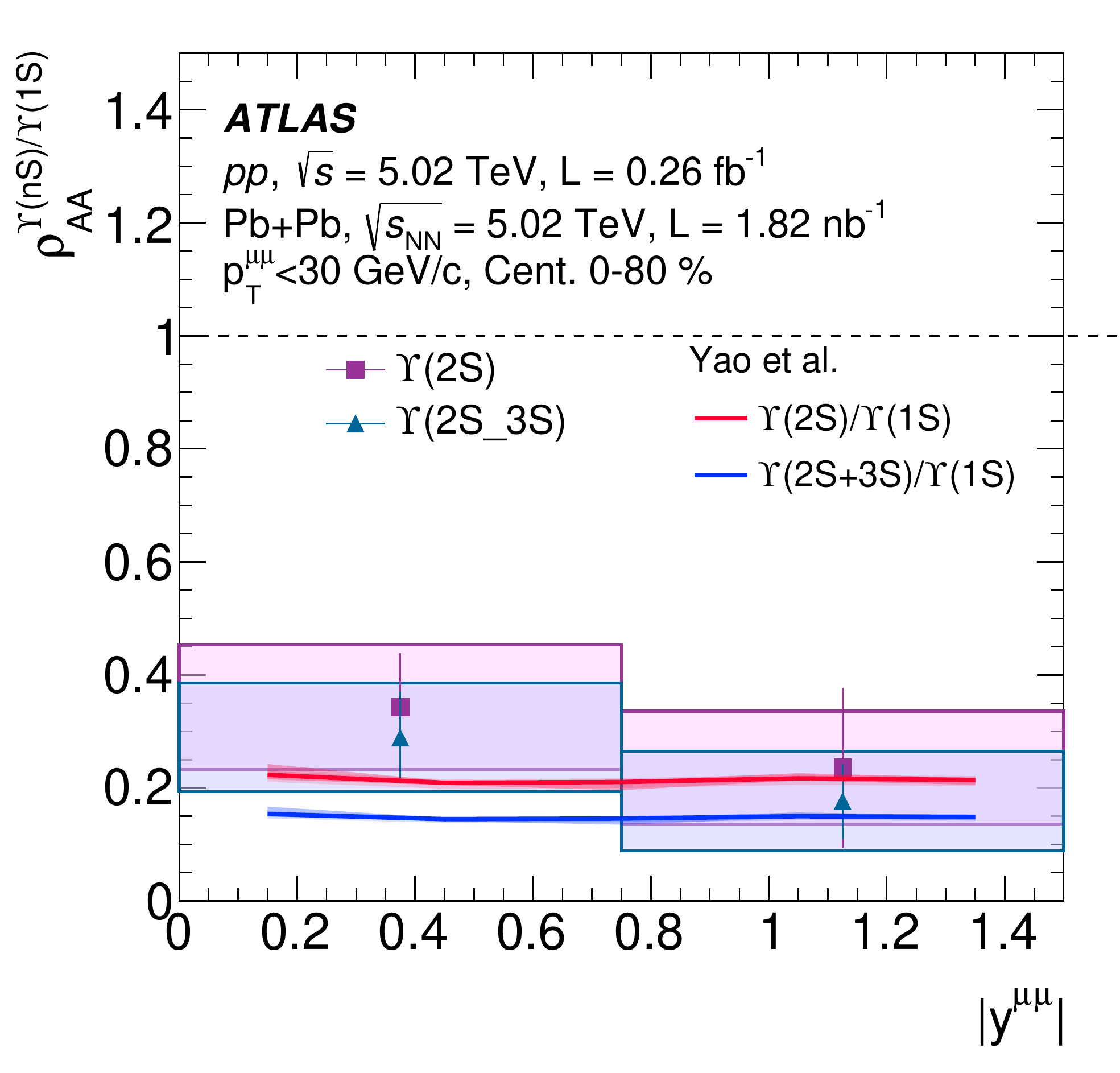}
\caption{The nuclear modification factor \RAA of $\Upsi$, $\Upsii$ and $\Upsiex$ (left column) and the double ratio $\rho_{\mathrm{AA}}^{\Upsins/\Upsi}$ for $\Upsii$ and $\Upsiex$ (right column) as functions of centrality (top row), $\pt^{\mu\mu}$ (center row) and rapidity (bottom row) at $5.02~\TeV$ per nucleon--nucleon pair compared to a calculation by X.~Yao et al., Ref.~\cite{Yao:2021} (solid curves). Dark color bands show theory uncertainty due to the nPDF choice. Light color bands show model uncertainties due to varying the calculation parameters by $\pm10\%$.}
\label{fig:yao_raa}
\end{center}
\end{figure}
 
All three models are in agreement with the data within experimental and
theoretical uncertainties.
It is notable that all three theoretical results were calculated after the
publication of a CMS measurement of $\varUpsilon$ suppression in Ref.~\cite{CMS-HIN-16-023} at the same beam energy.
 
\begin{figure}[htb]
\begin{center}
\includegraphics[width=\fourfigsize]{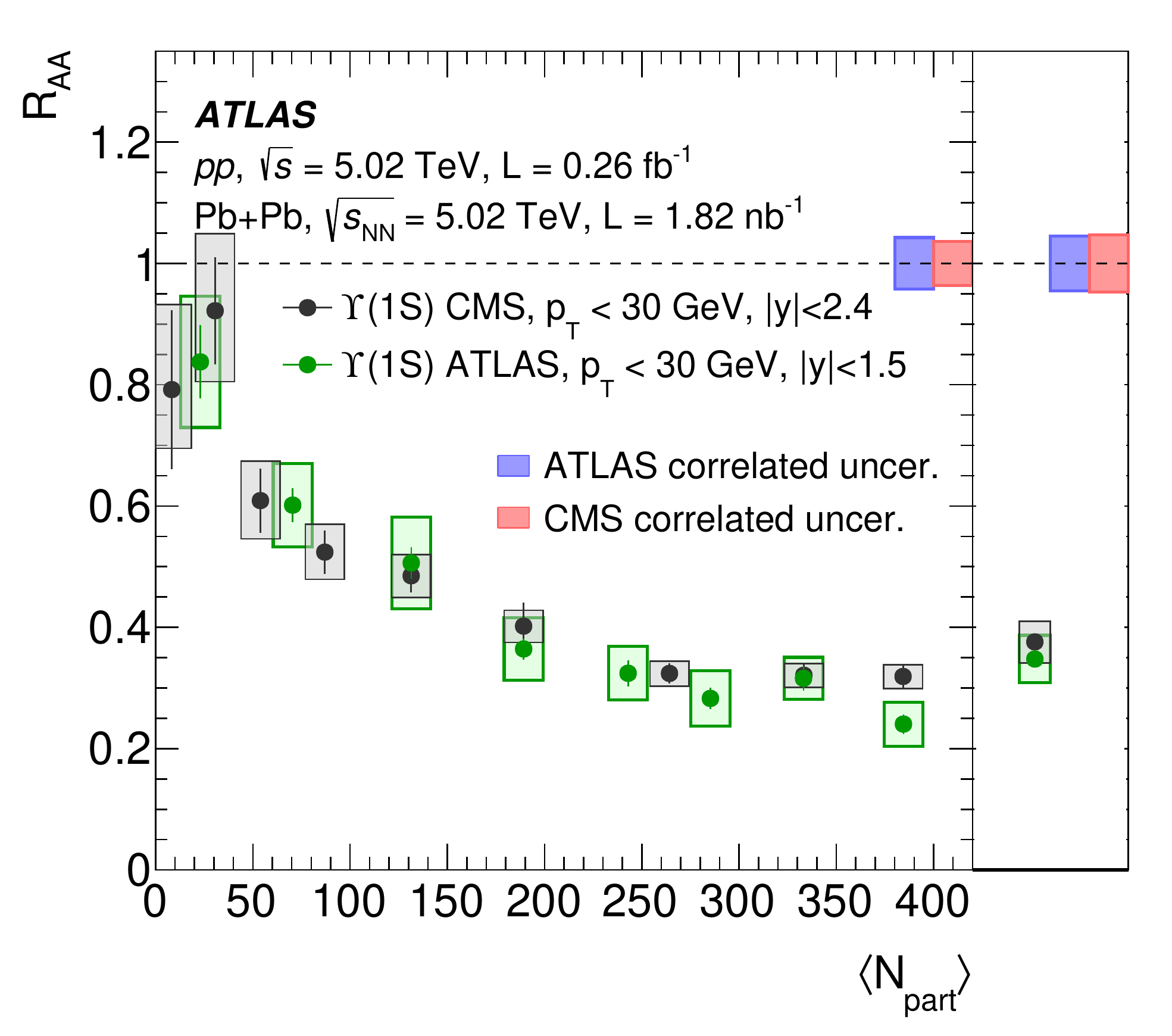}
\includegraphics[width=\fourfigsize]{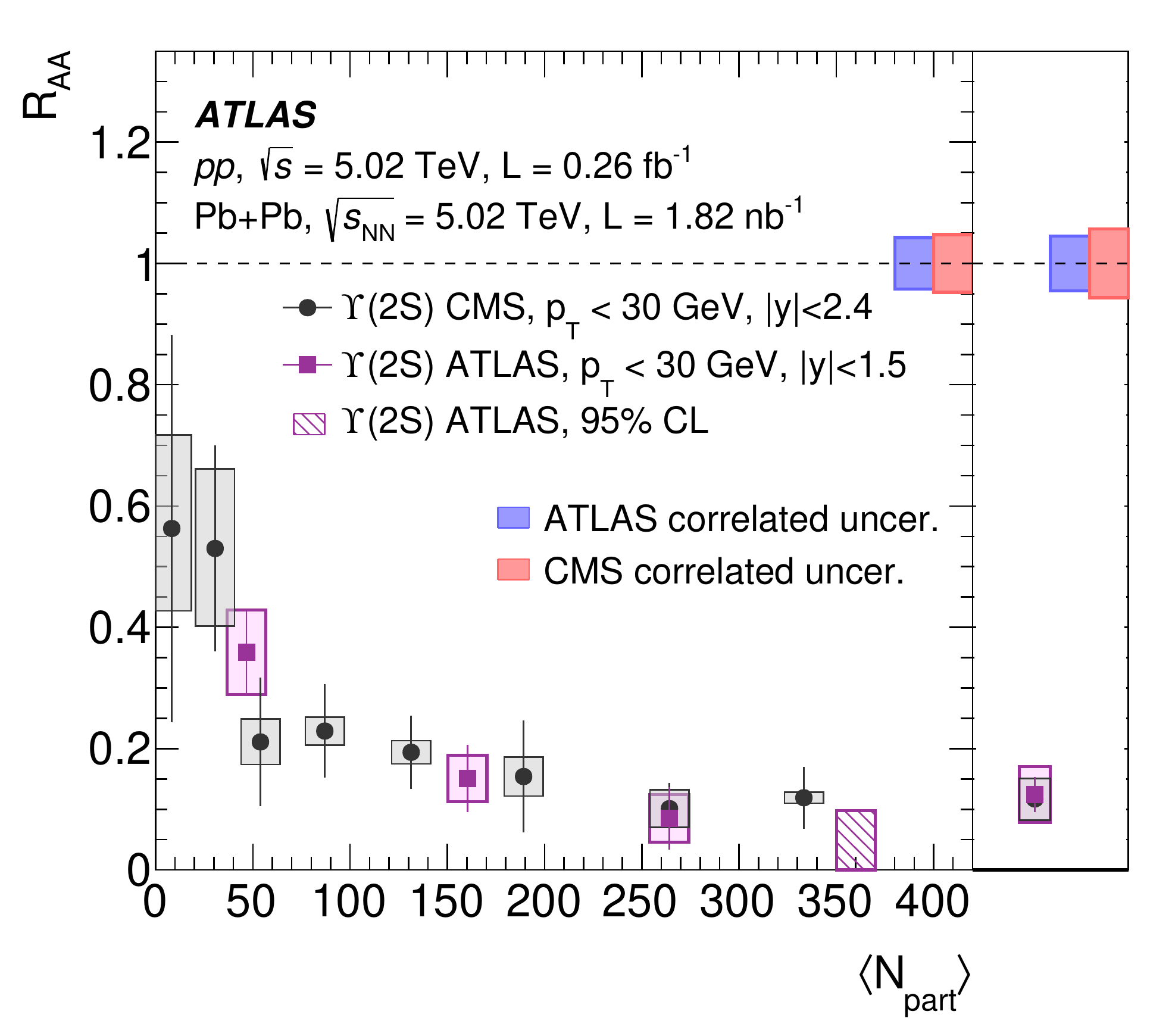}
\caption{The nuclear modification factor \RAA of $\Upsi$ (left) and  $\Upsii$ (right)
compared to the previous CMS measurement~\cite{CMS-HIN-16-008}.}
\label{fig:compcms}
\end{center}
\end{figure}

Figure~\ref{fig:compcms} shows the ATLAS results presented in this paper
compared with CMS published values~\cite{CMS-HIN-16-008}
for the \Upsi (left) and \Upsii (right) states.
These measured nuclear modification factors, as well as the double ratios,
are consistent between ATLAS and CMS within uncertainties.
The kinematic selections are similar, though ATLAS uses a narrow rapidity
window $|y|<1.5$ compared with $|y|<2.4$ in CMS.
Published results by the ALICE
Collaboration~\cite{ALICE-Ups-2014, ALICE-Ups-2021} are at more
forward rapidities, and thus no direct comparison is made.
The sequential melting of Upsilon states is thus confirmed by the
present measurement.
Suppression of the \Upsi observed at the level of approximately 0.25–0.30
in central Pb+Pb collisions, indicates suppression of the directly
produced \Upsi and not just suppression of higher states
feeding down to the \Upsi thus providing important information on the
highest temperatures achieved in the QGP.
 
More recently, CMS presented a measurement which includes the first
observation of the \Upsiii in Pb+Pb collisions at the LHC~\cite{CMS:2022rna}.
The \Upsiii is even more
suppressed than the \Upsi and \Upsii, further quantifying the sequential
melting of quarkonium states in QGP. The combined nuclear modification
factor for the two excited states \Upsiex presented in this paper is
consistent with the \Upsiii CMS measurement.

Reference~\cite{Mocsy2013} quotes dissociation temperatures in the
case of an isotropic QGP of < 192~\MeV, 228~\MeV, and 593~\MeV\ for the
\Upsiii, \Upsii, \Upsi states, respectively.
It is notable that the \Upsi has a substantial decay feed-down fraction
from \Upsii, \Upsiii, and $\chi_{b}(1S)$, $\chi_{b}(2S)$ states, and that only
about half of \Upsi are directly produced.
Thus, the suppression at the level
of approximately 0.25--0.30 for the \Upsi observed in central Pb+Pb
collisions indicates at least some suppression of
the directly produced \Upsi, which has important implications regarding
the temperatures achieved in the QGP.
 
\FloatBarrier

\section{Conclusions}
 
This paper presents a measurement of \Ups{n} yields in \pp and \PbPb data
at $5.02~\TeV$ per nucleon--nucleon collision.
The measurement uses data from \pp collisions collected in 2017 with a
total integrated luminosity of $0.26~\ifb$ and \PbPb collisions collected
in 2015 and 2018 with total integrated luminosities of $0.44~\inb$ and $1.38~\inb$
respectively, recorded by the ATLAS experiment at the LHC.
The \pp and \PbPb measurements are used to obtain the nuclear modification
factor for \Ups{1}, \Ups{2}, and \Upsiex,
as well as excited-state to ground-state double ratios of $R_{\mathrm{AA}}$ for \Ups{2} and
\Upsiex as functions
of transverse momentum, rapidity, and centrality. Both the \Ups{1} and \Ups{2} yields
are suppressed with increasing centrality in \PbPb compared to those in \pp
collisions, and the excited states are shown to be  stronger suppressed than the
ground state, resulting in double ratios smaller than one.
The \RAA value for \Upsiex appears to be systematically lower than that for
\Upsii, and so are the corresponding double ratios, which indicates that the \Upsiii
state is suppressed more than \Upsii.
 
The measured nuclear modification factors and double ratios are found to be
consistent with the previous CMS measurement~\cite{CMS-HIN-16-008}
at similar kinematical conditions.
 
All theoretical calculations considered in this paper describe the data well and
incorporate deconfinement as a key ingredient in the
suppression of the Upsilon yields. Further discrimination between the different
implementations of these deconfinement effects require additional precision data from
upcoming runs at RHIC and the LHC.

\section*{Acknowledgments}


We thank CERN for the very successful operation of the LHC, as well as the
support staff from our institutions without whom ATLAS could not be
operated efficiently.
 
We acknowledge the support of
ANPCyT, Argentina;
YerPhI, Armenia;
ARC, Australia;
BMWFW and FWF, Austria;
ANAS, Azerbaijan;
CNPq and FAPESP, Brazil;
NSERC, NRC and CFI, Canada;
CERN;
ANID, Chile;
CAS, MOST and NSFC, China;
Minciencias, Colombia;
MEYS CR, Czech Republic;
DNRF and DNSRC, Denmark;
IN2P3-CNRS and CEA-DRF/IRFU, France;
SRNSFG, Georgia;
BMBF, HGF and MPG, Germany;
GSRI, Greece;
RGC and Hong Kong SAR, China;
ISF and Benoziyo Center, Israel;
INFN, Italy;
MEXT and JSPS, Japan;
CNRST, Morocco;
NWO, Netherlands;
RCN, Norway;
MEiN, Poland;
FCT, Portugal;
MNE/IFA, Romania;
MESTD, Serbia;
MSSR, Slovakia;
ARRS and MIZ\v{S}, Slovenia;
DSI/NRF, South Africa;
MICINN, Spain;
SRC and Wallenberg Foundation, Sweden;
SERI, SNSF and Cantons of Bern and Geneva, Switzerland;
MOST, Taiwan;
TENMAK, T\"urkiye;
STFC, United Kingdom;
DOE and NSF, United States of America.
In addition, individual groups and members have received support from
BCKDF, CANARIE, Compute Canada and CRC, Canada;
PRIMUS 21/SCI/017 and UNCE SCI/013, Czech Republic;
COST, ERC, ERDF, Horizon 2020 and Marie Sk{\l}odowska-Curie Actions, European Union;
Investissements d'Avenir Labex, Investissements d'Avenir Idex and ANR, France;
DFG and AvH Foundation, Germany;
Herakleitos, Thales and Aristeia programmes co-financed by EU-ESF and the Greek NSRF, Greece;
BSF-NSF and MINERVA, Israel;
Norwegian Financial Mechanism 2014-2021, Norway;
NCN and NAWA, Poland;
La Caixa Banking Foundation, CERCA Programme Generalitat de Catalunya and PROMETEO and GenT Programmes Generalitat Valenciana, Spain;
G\"{o}ran Gustafssons Stiftelse, Sweden;
The Royal Society and Leverhulme Trust, United Kingdom.
 
The crucial computing support from all WLCG partners is acknowledged gratefully, in particular from CERN, the ATLAS Tier-1 facilities at TRIUMF (Canada), NDGF (Denmark, Norway, Sweden), CC-IN2P3 (France), KIT/GridKA (Germany), INFN-CNAF (Italy), NL-T1 (Netherlands), PIC (Spain), ASGC (Taiwan), RAL (UK) and BNL (USA), the Tier-2 facilities worldwide and large non-WLCG resource providers. Major contributors of computing resources are listed in Ref.~\cite{ATL-SOFT-PUB-2021-003}.


\printbibliography

@Article{     Strickland:2021,
  author        = {Brambilla, Nora and others},
  title         = {Bottomonium production in heavy-ion collisions using quantum trajectories: Differential observables and momentum anisotropy.},
    journal        = "Phys. Rev. D",
    volume         = "\textbf{104}",
    pages          = "094049",
  archiveprefix = {arXiv},
  eprint        = {2107.06222},
  primaryclass  = {hep-ph},
  doi           = {10.1103/PhysRevD.104.094049},
  year          = 2021,
  note          = { and private communication}
}

@Article{     Du:2017,
    author         = {X. Du and M. He, and R. Rapp},
    title          = {Color Screening and Regeneration of Bottomonia in High-Energy Heavy-Ion Collisions.},
    journal        = "Phys. Rev. C",
    volume         = "96",
    year           = "2017",
    pages          = "054901",
    doi            = "10.1103/PhysRevC.96.054901",
    eprint         = "1706.08670",
    archivePrefix  = "arXiv",
    primaryClass   = "hep-ph",
    note          = { and private communication}
}

@article{Matsui:1986dk,
      author         = "Matsui, T. and Satz, H.",
      title          = "{$J/\psi$ suppression by quark-gluon plasma formation}",
      journal        = "Phys. Lett.",
      volume         = "\textbf{B} 178",
      pages          = "416",
      doi            = "10.1016/0370-2693(86)91404-8",
      year           = "1986",
      reportNumber   = "BNL-38344",
      SLACcitation   = "%%CITATION = PHLTA,B178,416;%%",
}

@article{Karsch:1987pv,
      author         = "Karsch, F. and Mehr, M. T. and Satz, H.",
      title          = "{Color screening and deconfinement for bound states of
                        heavy quarks}",
      journal        = "Z. Phys.",
      volume         = "\textbf{C} 37",
      pages          = "617",
      doi            = "10.1007/BF01549722",
      year           = "1988",
      reportNumber   = "BNL-40122",
      SLACcitation   = "%%CITATION = ZEPYA,C37,617;%%",
}

@article{Digal:2001iu,
      author         = "Digal, S. and Petreczky, P. and Satz, H.",
      title          = "{String breaking and quarkonium dissociation at finite
                        temperatures}",
      journal        = "Phys. Lett.",
      volume         = "\textbf{B} 514",
      pages          = "57-62",
      doi            = "10.1016/S0370-2693(01)00803-6",
      year           = "2001",
      eprint         = "hep-ph/0105234",
      archivePrefix  = "arXiv",
      primaryClass   = "hep-ph",
      reportNumber   = "BI-TP-2001-08",
      SLACcitation   = "%%CITATION = HEP-PH/0105234;%%",
}

@article{Glauber,
      author         = {Miller, Michael L. and Reygers, Klaus and Sanders, Stephen J. and Steinberg, Peter},
      title          = {Glauber Modeling in High-Energy Nuclear Collisions},
      journal        = "Ann. Rev. Nucl. Part. Sci.",
      volume         = "57",
      pages          = "205",
      doi            = "10.1146/annurev.nucl.57.090506.123020",
      year           = "2007",
      eprint         = "0701025",
      archivePrefix  = "arXiv",
      primaryClass   = "nucl-ex",
}

@Article{     Yao:2021,
  author        = {Yao, X. and Ke, W. and Xu, Y. and Bass, S.~A. and Müller, B.},
  title         = {Coupled Boltzmann transport equations of heavy quarks and quarkonia in quark-gluon plasma.},
      journal        = "JHEP",
      volume         = "\textbf{2021}",
    pages          = "46",
  archiveprefix = {arXiv},
  eprint        = {2004.06746},
  primaryclass  = {hep-ph},
  doi           = {10.1007/JHEP01(2021)046},
  year          = 2021,
  note          = { and private communication}
}

@booklet{paper:CB, 
     author         = "Oreglia, M.", 
     title          = "{A Study of the Reactions $\psi^\prime \to \gamma\gamma\psi$}", 
     school         = "SLAC", 
     url            = "https://www.slac.stanford.edu/cgi-wrap/getdoc/slac-r-236.pdf", 
     year           = "1980", 
     reportNumber   = "SLAC-R-0236", 
     SLACcitation   = "%%CITATION = SLAC-0236;%%" 
}

@Article{NRQCD:1995,
      author         	= "Bodwin, G. T. and Braaten, E. and Lepage, G. P.",
      title          	= "{Rigorous QCD analysis of inclusive annihilation and production of heavy quarkonium}",
      journal       	= "Phys. Rev. D",
      volume         	= "51",
      pages          	= "1125-1171",
      year           	= "1995",
      eprint         	= "hep-ph/9407339",
      archivePrefix  	= "arXiv",
      doi               = "https://doi.org/10.48550/arXiv.hep-ph/9407339"
}

@article{LHCb_Jpsi_polarization,
	Archiveprefix = {arXiv},
	Author = {{LHCb Collaboration}},
	Collaboration = {{LHCb}},
	Doi = {10.1140/epjc/s10052-013-2631-3},
	Eprint = {1307.6379},
	Journal = {Eur. Phys. J. C},
	Pages = {2631},
	Primaryclass = {hep-ex},
	Title = {{Measurement of $J/\psi$ polarization in pp collisions at $\sqrt{s} = 7~\mathrm{TeV}$}},
	Volume = {73},
	Year = {2013}
}

@Article{ALICE-Jpsi-2017,
    author         = "{ALICE Collaboration}",
    title          = "{Differential studies of inclusive $J/\psi$ and $\psi(2S)$ production at forward rapidity in Pb–Pb collisions at $\sqrt{s_{NN}}=2.76$ TeV.}",
    journal        = "JHEP",
    volume         = "05",
    year           = "2016",
    pages          = "179",
    doi            = "10.1007/JHEP05(2016)179",
    eprint         = "1506.08804",
    archivePrefix  = "arXiv",
    primaryClass   = "nucl-ex",
}

@Article{STAR-Jpsi-2017,
    author         = "{STAR Collaboration}",
    collaboration  = "{STAR Collaboration}",
    title          = "{Energy dependence of $J\psi$ production in Au+Au collisions at $\sqrt{s_{NN}}$ = 39, 62.4 and 200 GeV.}",
    journal        = "Phys. Lett. B",
    volume         = "771",
    year           = "2017",
    pages          = "13",
    doi            = "https://doi.org/10.1016/j.physletb.2017.04.078",
    eprint         = "arXiv:1607.07517",
    archivePrefix  = "arXiv",
}

@Article{PHENIX-Jpsi-2007,
    author         = "{PHENIX Collaboration}",
    collaboration  = "{PHENIX Collaboration}",
    title          = "{$J/\psi$ production vs centrality, transverse momentum, and rapidity in Au+Au collisions at $\sqrt{s_{NN}}=200$ GeV.}",
    journal        = "Phys. Rev. Lett.",
    volume         = "98",
    year           = "2007",
    pages          = "232301",
    doi            = "https://doi.org/10.1103/PhysRevLett.98.232301",
}

@Article{PHENIX-Ups-2015,
    author         = "{PHENIX Collaboration}",
    title          = "{Measurement of $\varUpsilon(1S+2S+3S)$ production in p+p and Au+Au collisions at $\sqrt{s_{NN}}=200$ GeV.}",
    journal        = "Phys. Rev. \textbf{C}",
    volume         = "91",
    year           = "2015",
    pages          = "024913",
    doi            = "10.1103/PhysRevC.91.024913",
    eprint         = "1404.2246",
    archivePrefix  = "arXiv",   
}

@Article{ALICE-Ups-2014,
    author         = "{ALICE Collaboration}",
    title          = "{Suppression of $\varUpsilon(1S)$ at forward rapidity in Pb–Pb collisions at $\sqrt{s_{NN}}=2.76$ TeV}",
    journal        = "Phys. Lett. B",
    volume         = "738",
    year           = "2014",
    pages          = "361",
    doi            = "10.1016/j.physletb.2014.10.001",
    eprint         = "arXiv:1405.4493",
    archivePrefix  = "arXiv",
}

@Article{ALICE-Ups-2021,
    author         = "{ALICE Collaboration}",
    title          = "{$\varUpsilon(1S)$ production and nuclear modification at forward rapidity in Pb–Pb collisions at $\sqrt{s_{NN}}=5.02$ TeV}",
    journal        = "Phys. Lett. B",
    volume         = "822",
    year           = "2021",
    pages          = "136579",
    doi            = "10.1016/j.physletb.2021.136579",
    eprint         = "arXiv:2011.05758",
    archivePrefix  = "arXiv",
}

@Article{STAR-Ups-2014,
    author         = "{STAR Collaboration}",
    title          = "{Suppression of $\varUpsilon$ production in d+Au and Au+Au collisions at $\sqrt{s_{NN}}=200$ GeV.}",
    journal        = "Phys. Lett. B",
    volume         = "735",
    year           = "2014",
    pages          = "127",
    doi            = "https://doi.org/10.1016/j.physletb.2014.06.028",
    eprint         = "1312.3675",
    archivePrefix  = "arXiv",
}

@Article{PDG-ptep,
      author         = "{Particle Data Group}",
      title          = "{Review of Particle Physics}",
      collaboration  = "Particle Data Group",
      journal        = "Prog. Theor. Exp. Phys.",
      volume         = "2020",
      pages          = "083C01",
      year           = "2020",
      doi            = "https://doi.org/10.1093/ptep/ptaa104",
}

@Article{THRev1,
      author         	= "Brambilla, N. and others",
      title          	= "{Heavy quarkonium: progress, puzzles, and opportunities.}",
      journal       	= "Eur. Phys. J. C",
      volume         	= "71",
      pages          	= "1534",
      year           	= "2011",
      eprint         	= "1010.5827",
      archivePrefix  	= "arXiv",
      doi               = "10.1140/epjc/s10052-010-1534-9",
}

@Article{THRev2,
      author         	= "Andronic, A. and others",
      title          	= "{Heavy-flavour and quarkonium production in the LHC era: from proton–proton to heavy-ion collisions.}",
      journal       	= "Eur. Phys. J. C",
      volume         	= "76",
      pages          	= "107",
      year           	= "2016",
      eprint         	= "1506.03981",
      archivePrefix  	= "arXiv",
      doi               = "10.1140/epjc/s10052-015-3819-5",
}

@Article{DuRapp,
      author         	= "Rapp, R. and Du, X.",
      title          	= "{Theoretical Perspective on Quarkonia from SPS via RHIC to LHC.}",
      journal       	= "Nucl. Phys. A",
      volume         	= "967",
      pages          	= "216-224",
      year           	= "2027",
      eprint         	= "",
      archivePrefix  	= "arXiv",
      doi               = "https://doi.org/10.1016/j.nuclphysa.2017.05.097",
}

@Article{Greiner2018,
      author         	= "N. Krenz, and H. van Hees, and C. Greiner",
      title          	= "{Quarkonia production in a Langevin approach.}",
      journal       	= "J. Phys. Conf. Ser.",
      volume         	= "1070",
      pages          	= "012008",
      year           	= "2018",
      eprint         	= "1805.10798",
      archivePrefix  	= "arXiv",
      doi               = "10.1088/1742-6596/1070/1/012008",
}

@Article{Mocsy2013,
      author         	= "A. Mocsy and P. Petreczky and M. Strickland",
      title          	= "{Quarkonia in the Quark Gluon Plasma.}",
      journal       	= "Int. J. Mod. Phys. A",
      volume         	= "28",
      pages          	= "1340012",
      year           	= "2013",
      eprint         	= "1302.2180",
      archivePrefix  	= "arXiv",
      doi               = "10.1142/S0217751X13400125",
}

@Article{SPS2,
      author         	= "{NA50 Collaboration}",
      title          	= "{$\psi^{\prime}$ production in Pb-Pb collisions at 158 GeV/nucleon}",
      journal       	= "Eur. Phys. J. C",
      volume         	= "49",
      pages          	= "559-567",
      year           	= "2007",
      eprint         	= "nucl-ex/0612013",
      archivePrefix  	= "arXiv",
      doi               = "10.1140/epjc/s10052-006-0153-y",
}

@Article{SPS1,
      author         	= "{NA50 Collaboration}",
      title          	= "{A new measurement of $J/\psi$ suppression in Pb-Pb collisions at 158 GeV per nucleon}",
      journal       	= "Eur. Phys. J. C",
      volume         	= "39",
      pages          	= "335-345",
      year           	= "2005",
      eprint         	= "hep-ex/0412036",
      archivePrefix  	= "arXiv",
      doi               = "https://doi.org/10.1140/epjc/s2004-02107-9",
}

@Booklet{CMS:2022rna,
    author = "{CMS Collaboration}",
    title = "{Observation of the $\varUpsilon\textrm{(3S)}$ meson and sequential suppression of $\varUpsilon$ states in PbPb collisions at $\sqrt{\mathrm{s_{NN}}}=5.02~\mathrm{TeV}$}",
    howpublished   = "{CMS-PAS-HIN-21-007}",
    url            = "https://inspirehep.net/literature/2071548",
    year = "2022"
}

@Article{STAR-UU,
    author         = "{STAR Collaboration}",
    title          = "{Upsilon production in U+U collisions at 193 GeV with the STAR experiment}",
    journal        = "Phys. Rev. C",
    volume         = "94",
    year           = "2016",
    pages          = "064904",
    doi            = "https://doi.org/10.1103/PhysRevC.94.064904",
    eprint         = "1608.06487",
    archivePrefix  = "arXiv",
}

@Article{STAR-NEW,
    author         = "{STAR Collaboration}",
    title          = "{Observation of sequential $\varUpsilon$ suppression in Au+Au collisions at $\sqrt{\mathrm{s_{NN}}}=200~\mathrm{GeV}$ with the STAR experiment}",
    year           = "2022",
    doi            = "https://doi.org/10.48550/arXiv.2207.06568",
    eprint         = "2207.06568",
    archivePrefix  = "arXiv",
}

@Article{Pumplin:2002vw,
     author    = "Pumplin, J. and others",
     title     = "{New Generation of Parton Distributions with Uncertainties from Global QCD Analysis}",
     journal   = "JHEP",
     volume    = "07",
     year      = "2002",
     pages     = "012",
     doi       = "10.1088/1126-6708/2002/07/012",
     eprint    = "hep-ph/0201195",
     archivePrefix = "arXiv",
     SLACcitation  = "%%CITATION = HEP-PH/0201195;%%"
}

@article{Sjostrand:2014zea,
      author         = "Sj{\"o}strand, Torbj{\"o}rn and Ask, Stefan and Christiansen,
                        Jesper R. and Corke, Richard and Desai, Nishita and Ilten,
                        Philip and Mrenna, Stephen and Prestel, Stefan and
                        Rasmussen, Christine O. and Skands, Peter Z.",
      title          = "{An introduction to PYTHIA 8.2}",
      journal        = "Comput. Phys. Commun.",
      volume         = "191",
      year           = "2015",
      pages          = "159",
      doi            = "10.1016/j.cpc.2015.01.024",
      eprint         = "1410.3012",
      archivePrefix  = "arXiv",
      primaryClass   = "hep-ph",
      reportNumber   = "LU-TP-14-36, MCNET-14-22, CERN-PH-TH-2014-190,
                        FERMILAB-PUB-14-316-CD, DESY-14-178, SLAC-PUB-16122,
                        --FERMILAB-PUB-14-316-CD",
      SLACcitation   = "%%CITATION = ARXIV:1410.3012;%%"
}

@article{Agostinelli:2002hh,
      author         = "Agostinelli, S. and others",
      title          = "{\textsc{Geant4} -- a simulation toolkit}",
      journal        = "Nucl. Instrum. Meth. A",
      volume         = "506",
      year           = "2003",
      pages          = "250",
      doi            = "10.1016/S0168-9002(03)01368-8",
      reportNumber   = "SLAC-PUB-9350, FERMILAB-PUB-03-339",
      SLACcitation   = "%%CITATION = NUIMA,A506,250;%%"
}

@Article{PERF-2007-01,
    author         = "{ATLAS Collaboration}",
    title          = "{The ATLAS Experiment at the CERN Large Hadron Collider}",
    journal        = "JINST",
    volume         = "3",
    year           = "2008",
    pages          = "S08003",
    doi            = "10.1088/1748-0221/3/08/S08003",
    primaryClass   = "hep-ex",
}

@Article{SOFT-2010-01,
    author         = "{ATLAS Collaboration}",
    title          = "{The ATLAS Simulation Infrastructure}",
    journal        = "Eur. Phys. J. C",
    volume         = "70",
    year           = "2010",
    pages          = "823",
    doi            = "10.1140/epjc/s10052-010-1429-9",
    eprint         = "1005.4568",
    archivePrefix  = "arXiv",
    primaryClass   = "physics.ins-det",
}

@Article{BPHY-2011-01,
    author         = "{ATLAS Collaboration}",
    title          = "{Measurement of the \(\Upsilon(1\text{S})\) Production Cross-Section in \(pp\) Collisions at \(\sqrt{s} = 7\,\text{TeV}\) in ATLAS}",
    journal        = "Phys. Lett. B",
    volume         = "705",
    year           = "2011",
    pages          = "9",
    doi            = "10.1016/j.physletb.2011.09.092",
    reportNumber   = "CERN-PH-EP-2011-083",
    eprint         = "1106.5325",
    archivePrefix  = "arXiv",
    primaryClass   = "hep-ex",
}

@Article{HION-2014-05,
    author         = "{ATLAS Collaboration}",
    title          = "{Measurement of quarkonium production in proton--lead and proton--proton collisions at \(5.02\,\text{TeV}\) with the ATLAS detector}",
    journal        = "Eur. Phys. J. C",
    volume         = "78",
    year           = "2018",
    pages          = "171",
    doi            = "10.1140/epjc/s10052-018-5624-4",
    reportNumber   = "CERN-EP-2017-197",
    eprint         = "1709.03089",
    archivePrefix  = "arXiv",
    primaryClass   = "hep-ex",
}

@Article{PERF-2015-10,
    author         = "{ATLAS Collaboration}",
    title          = "{Muon reconstruction performance of the ATLAS detector in proton--proton collision data at \(\sqrt{s} = 13\,\text{TeV}\)}",
    journal        = "Eur. Phys. J. C",
    volume         = "76",
    year           = "2016",
    pages          = "292",
    doi            = "10.1140/epjc/s10052-016-4120-y",
    reportNumber   = "CERN-EP-2016-033",
    eprint         = "1603.05598",
    archivePrefix  = "arXiv",
    primaryClass   = "hep-ex",
}

@Article{HION-2016-06,
    author         = "{ATLAS Collaboration}",
    title          = "{Measurement of the azimuthal anisotropy of charged particles produced in \(\sqrt{s_{\text{NN}}} = 5.02\,\text{TeV}\) Pb+Pb collisions with the ATLAS detector}",
    journal        = "Eur. Phys. J. C",
    volume         = "78",
    year           = "2018",
    pages          = "997",
    doi            = "10.1140/epjc/s10052-018-6468-7",
    reportNumber   = "CERN-EP-2018-194",
    eprint         = "1808.03951",
    archivePrefix  = "arXiv",
    primaryClass   = "hep-ex",
}

@Article{HION-2016-07,
    author         = "{ATLAS Collaboration}",
    title          = "{Prompt and non-prompt \(J/\psi\) and \(\psi(2\text{S})\) suppression at high transverse momentum in \(5.02\,\text{TeV}\) Pb+Pb collisions with the ATLAS experiment}",
    journal        = "Eur. Phys. J. C",
    volume         = "78",
    year           = "2018",
    pages          = "762",
    doi            = "10.1140/epjc/s10052-018-6219-9",
    reportNumber   = "CERN-EP-2018-049",
    eprint         = "1805.04077",
    archivePrefix  = "arXiv",
    primaryClass   = "hep-ex",
}

@Article{TRIG-2016-01,
    author         = "{ATLAS Collaboration}",
    title          = "{Performance of the ATLAS trigger system in 2015}",
    journal        = "Eur. Phys. J. C",
    volume         = "77",
    year           = "2017",
    pages          = "317",
    doi            = "10.1140/epjc/s10052-017-4852-3",
    reportNumber   = "CERN-EP-2016-241",
    eprint         = "1611.09661",
    archivePrefix  = "arXiv",
    primaryClass   = "hep-ex",
}

@Article{PIX-2018-001,
    author         = "Abbott, B. and others",
    title          = "{Production and integration of the ATLAS Insertable B-Layer}",
    journal        = "JINST",
    volume         = "13",
    year           = "2018",
    pages          = "T05008",
    doi            = "10.1088/1748-0221/13/05/T05008",
    eprint         = "1803.00844",
    archivePrefix  = "arXiv",
    primaryClass   = "physics.ins-det",
}

@Article{HION-2021-12,
    author         = "{ATLAS Collaboration}",
    title          = "{Production of \(\varUpsilon(\text{nS})\) mesons in Pb+Pb and \(pp\) collisions at \(5.02\,\text{TeV}\)}",
    journal        = "Phys. Rev. C",
    volume         = "107",
    year           = "2022",
    pages          = "054912",
    doi            = "10.1103/PhysRevC.107.054912",
    reportNumber   = "CERN-EP-2022-045",
    eprint         = "2205.03042",
    archivePrefix  = "arXiv",
    primaryClass   = "nucl-ex",
}

@Booklet{ATL-SOFT-PUB-2021-001,
    author         = "{ATLAS Collaboration}",
    title          = "{The ATLAS Collaboration Software and Firmware}",
    howpublished   = "{ATL-SOFT-PUB-2021-001}",
    url            = "https://cds.cern.ch/record/2767187",
    year           = "2021",
}

@Booklet{ATL-SOFT-PUB-2021-003,
    author         = "{ATLAS Collaboration}",
    title          = "{ATLAS Computing Acknowledgements}",
    howpublished   = "{ATL-SOFT-PUB-2021-003}",
    url            = "https://cds.cern.ch/record/2776662",
    year           = "2021",
}

@Report{ATLAS-TDR-19,
    author         = "{ATLAS Collaboration}",
    title          = "{ATLAS Insertable B-Layer: Technical Design Report}",
    type           = "ATLAS-TDR-19; CERN-LHCC-2010-013",
    year           = "2010",
    url            = "https://cds.cern.ch/record/1291633",
    related        = "ATLAS-TDR-19-addm",
    relatedstring  = "Addendum:",
}

@Article{CMS-BPH-11-023,
    author         = "{CMS Collaboration}",
    title          = "{Measurement of the \(\Upsilon(1\text{S})\), \(\Upsilon(2\text{S})\) and \(\Upsilon(3\text{S})\) Polarizations in \(pp\) Collisions at \(\sqrt{s} = 7\,\text{TeV}\)}",
    journal        = "Phys. Rev. Lett.",
    volume         = "110",
    year           = "2013",
    pages          = "081802",
    doi            = "10.1103/PhysRevLett.110.081802",
    reportNumber   = "CERN-PH-EP-2012-205",
    eprint         = "1209.2922",
    archivePrefix  = "arXiv",
    primaryClass   = "hep-ex",
}

@Article{CMS-BPH-13-003,
    author         = "{CMS Collaboration}",
    title          = "{Measurement of the prompt \(J/\psi\) and \(\psi(2\text{S})\) polarizations in \(pp\) collisions at \(\sqrt{s} = 7\,\text{TeV}\)}",
    journal        = "Phys. Lett. B",
    volume         = "727",
    year           = "2013",
    pages          = "381",
    doi            = "10.1016/j.physletb.2013.10.055",
    reportNumber   = "CERN-PH-EP-2013-140",
    eprint         = "1307.6070",
    archivePrefix  = "arXiv",
    primaryClass   = "hep-ex",
}

@Article{CMS-HIN-14-005,
    author         = "{CMS Collaboration}",
    title          = "{Suppression and azimuthal anisotropy of prompt and nonprompt \(J/\psi\) production in PbPb collisions at \(\sqrt{s_{\text{NN}}} = 2.76\,\text{TeV}\)}",
    journal        = "Eur. Phys. J. C",
    volume         = "77",
    year           = "2017",
    pages          = "252",
    doi            = "10.1140/epjc/s10052-017-4781-1",
    reportNumber   = "CERN-EP-2016-243",
    eprint         = "1610.00613",
    archivePrefix  = "arXiv",
    primaryClass   = "hep-ex",
}

@Article{CMS-HIN-16-008,
    author         = "{CMS Collaboration}",
    title          = "{Suppression of excited \(\Upsilon\) States Relative to the Ground State in PbPb Collisions at \(\sqrt{s_{\text{NN}}} = 5.02\,\text{TeV}\)}",
    journal        = "Phys. Rev. Lett.",
    volume         = "120",
    year           = "2018",
    pages          = "142301",
    doi            = "10.1103/PhysRevLett.120.142301",
    reportNumber   = "CERN-EP-2017-080",
    eprint         = "1706.05984",
    archivePrefix  = "arXiv",
    primaryClass   = "hep-ex",
}

@Article{CMS-HIN-16-023,
    author         = "{CMS Collaboration}",
    title          = "{Measurement of nuclear modification factors of \(\Upsilon(1\text{S})\), \(\Upsilon(2\text{S})\), and \(\Upsilon(3\text{S})\) mesons in PbPb collisions at \(\sqrt{s_{\text{NN}}} = 5.02\,\text{TeV}\)}",
    journal        = "Phys. Lett. B",
    volume         = "790",
    year           = "2019",
    pages          = "270",
    doi            = "10.1016/j.physletb.2019.01.006",
    reportNumber   = "CERN-EP-2018-110",
    eprint         = "1805.09215",
    archivePrefix  = "arXiv",
    primaryClass   = "hep-ex",
}

@Booklet{ATLAS-CONF-2019-021,
    author         = "{ATLAS Collaboration}",
    title          = "{Luminosity determination in \(pp\) collisions at \(\sqrt{s} = 13\,\text{TeV}\) using the ATLAS detector at the LHC}",
    howpublished   = "{ATLAS-CONF-2019-021}",
    url            = "https://cds.cern.ch/record/2677054",
    year           = "2019",
}

@Booklet{ATL-PHYS-PUB-2010-013,
    author         = "{ATLAS Collaboration}",
    title          = "{The simulation principle and performance of the ATLAS fast calorimeter simulation FastCaloSim}",
    howpublished   = "{ATL-PHYS-PUB-2010-013}",
    url            = "https://cds.cern.ch/record/1300517",
    year           = "2010",
}
 
 
\begin{flushleft}
\hypersetup{urlcolor=black}
{\Large The ATLAS Collaboration}

\bigskip

\AtlasOrcid[0000-0002-6665-4934]{G.~Aad}$^\textrm{\scriptsize 101}$,
\AtlasOrcid[0000-0002-5888-2734]{B.~Abbott}$^\textrm{\scriptsize 119}$,
\AtlasOrcid[0000-0002-7248-3203]{D.C.~Abbott}$^\textrm{\scriptsize 102}$,
\AtlasOrcid[0000-0002-1002-1652]{K.~Abeling}$^\textrm{\scriptsize 55}$,
\AtlasOrcid[0000-0002-8496-9294]{S.H.~Abidi}$^\textrm{\scriptsize 29}$,
\AtlasOrcid[0000-0002-9987-2292]{A.~Aboulhorma}$^\textrm{\scriptsize 35e}$,
\AtlasOrcid[0000-0001-5329-6640]{H.~Abramowicz}$^\textrm{\scriptsize 150}$,
\AtlasOrcid[0000-0002-1599-2896]{H.~Abreu}$^\textrm{\scriptsize 149}$,
\AtlasOrcid[0000-0003-0403-3697]{Y.~Abulaiti}$^\textrm{\scriptsize 116}$,
\AtlasOrcid[0000-0003-0762-7204]{A.C.~Abusleme~Hoffman}$^\textrm{\scriptsize 136a}$
% ...
\bigskip
\\

% ...
$^{29}$Physics Department, Brookhaven National Laboratory, Upton NY; United States of America.\\
% ...
$^{35}$$^{(a)}$Facult\'e des Sciences Ain Chock, R\'eseau Universitaire de Physique des Hautes Energies - Universit\'e Hassan II, Casablanca;$^{(b)}$Facult\'{e} des Sciences, Universit\'{e} Ibn-Tofail, K\'{e}nitra;$^{(c)}$Facult\'e des Sciences Semlalia, Universit\'e Cadi Ayyad, LPHEA-Marrakech;$^{(d)}$LPMR, Facult\'e des Sciences, Universit\'e Mohamed Premier, Oujda;$^{(e)}$Facult\'e des sciences, Universit\'e Mohammed V, Rabat;$^{(f)}$Institute of Applied Physics, Mohammed VI Polytechnic University, Ben Guerir; Morocco.\\
% ...
$^{55}$II. Physikalisches Institut, Georg-August-Universit\"{a}t G\"ottingen, G\"ottingen; Germany.\\
% ...
$^{101}$CPPM, Aix-Marseille Universit\'e, CNRS/IN2P3, Marseille; France.\\
$^{102}$Department of Physics, University of Massachusetts, Amherst MA; United States of America.\\
% ...
$^{116}$Department of Physics, New York University, New York NY; United States of America.\\
% ...
$^{119}$Homer L. Dodge Department of Physics and Astronomy, University of Oklahoma, Norman OK; United States of America.\\
% ...
$^{136}$$^{(a)}$Departamento de F\'isica, Pontificia Universidad Cat\'olica de Chile, Santiago;$^{(b)}$Millennium Institute for Subatomic physics at high energy frontier (SAPHIR), Santiago;$^{(c)}$Instituto de Investigaci\'on Multidisciplinario en Ciencia y Tecnolog\'ia, y Departamento de F\'isica, Universidad de La Serena;$^{(d)}$Universidad Andres Bello, Department of Physics, Santiago;$^{(e)}$Instituto de Alta Investigaci\'on, Universidad de Tarapac\'a, Arica;$^{(f)}$Departamento de F\'isica, Universidad T\'ecnica Federico Santa Mar\'ia, Valpara\'iso; Chile.\\
% ...
$^{149}$Department of Physics, Technion, Israel Institute of Technology, Haifa; Israel.\\
$^{150}$Raymond and Beverly Sackler School of Physics and Astronomy, Tel Aviv University, Tel Aviv; Israel.\\
% ...
$^{a}$ Also Affiliated with an institute covered by a cooperation agreement with CERN.\\
$^{b}$ Also at Borough of Manhattan Community College, City University of New York, New York NY; United States of America.\\
$^{c}$ Also at Bruno Kessler Foundation, Trento; Italy.\\
$^{d}$ Also at Center for High Energy Physics, Peking University; China.\\
$^{e}$ Also at Centro Studi e Ricerche Enrico Fermi; Italy.\\
$^{f}$ Also at CERN, Geneva; Switzerland.\\
% ...

\end{flushleft}


\end{document}